\documentclass[prb,superscriptaddress,twocolumn,showpacs,amsmath,amssymb]{revtex4}
\usepackage{graphicx} % Include figure files
\usepackage{bm}
\usepackage{txfonts}
\usepackage{mathrsfs}
\def\bk{{\mathbf k}}

\def\bp{{\mathbf p}}
\def\bq{{\mathbf q}}
\def\br{{\mathbf r}}

\def\b0{{\mathbf 0}}

\def\cS{{\cal S}}

\def\bra{\langle}
\def\ket{\rangle}
\def\up{\uparrow}
\def\down{\downarrow}

\def\alf{\alpha}
\def\eps{\epsilon}
\def\gam{\gamma}
\def\Gam{\Gamma}

\def\Lam{\Lambda}
\def\om{\omega}

\def\sg{\sigma}
\def\Sg{\Sigma}

\def\psib{\bar\psi}

\def\sgn{{\rm sgn}}

\usepackage[]{psfrag}

\psfrag{alpha}{$\alpha$}
\psfrag{gap}{$\Delta$}
\psfrag{lambda}{$\lambda$}
\psfrag{L}{$\Lambda$}
\psfrag{L_crit}{$\Lambda_c$}
\psfrag{alphgap}{$\alpha\,,\,\,\Delta$}
\psfrag{Lc_alph_gap}{$\Lambda_{c}\,,\,\,\alpha\,,\,\,\Delta$}

\psfrag{m_bosonic}{$m^{2}_{b},\,\,m^{2}_{\sigma}$}
\psfrag{Z_factors}{$Z_{\pi},\,\,Z_{\sigma}$}
\psfrag{g_factors}{$g_{\pi},\,\,g_{\sigma}$}

\begin{document}

\title{Low energy singularities in the ground state of fermionic superfluids}
\author{B. Obert}
\affiliation{Max-Planck-Institute for Solid State Research,
 Heisenbergstr.\ 1, D-70569 Stuttgart, Germany} 
\author{C. Husemann}
\altaffiliation{Present address: Carl Zeiss AG, Carl Zeiss Promenade 10,
 D-07745 Jena, Germany}
\affiliation{Max-Planck-Institute for Solid State Research,
 Heisenbergstr.\ 1, D-70569 Stuttgart, Germany}
\author{W. Metzner}
\affiliation{Max-Planck-Institute for Solid State Research,
 Heisenbergstr.\ 1, D-70569 Stuttgart, Germany} 
\date{\today}
\begin{abstract}
We analyze the effects of order parameter fluctuations on the ground
state of fully gapped charge-neutral fermionic superfluids. 
The Goldstone mode associated with the spontaneously broken symmetry
leads to a problem of coupled singularities in $d \leq 3$ dimensions.
We derive a minimal set of one-loop renormalization group equations 
which fully captures the interplay of the singularities.
The flow equations are based on a symmetry conserving truncation
of a scale dependent effective action.
We compute the low energy behavior of longitudinal, transverse 
and mixed order parameter correlations, and their impact on the 
fermionic gap.
We demonstrate analytically that cancellations protecting the 
Goldstone mode are respected by the flow, and we present a numerical
solution of the flow equations for the two-dimensional attractive
Hubbard model.
\end{abstract}

\maketitle

%%%%%%%%%%%%%%%%%%%%%%%%%%%%%%%%%%%%%%%%%%%%%%%%%%%%%%%%%%%%%%%%%%%%%%%%%%

\section{Introduction}

Spontaneously broken continuous symmetries in condensed matter and
high energy physics give rise to emergent collective excitations, 
well known as Goldstone modes.
Superfluidity in interacting Fermi and Bose systems is associated 
with a spontaneously broken $U(1)$ symmetry corresponding to particle
number conservation.
In charged superfluids the Goldstone modes are gapped by the coupling
to the electromagnetic field via the Anderson-Higgs mechanism.
\cite{andersen58,higgs64}
In superfluids made from charge neutral constituents such as superfluid
liquid Helium or ultracold atomic gases, the Goldstone modes are 
gapless and therefore have a significant impact on the low-energy 
behavior.

While the Goldstone modes themselves are protected by symmetry, they 
strongly affect {\em longitudinal} order parameter fluctuations, leading 
to drastic deviations from mean-field theory in dimensions $d \leq 3$
even at zero temperature.
This effect was intensively studied for the interacting Bose gas,
with important contributions scattered over several decades.
\cite{nepomnyashchy92,pistolesi04}
The Goldstone modes lead to infrared divergences in perturbation
theory. These divergences partially cancel each other due to symmetries,
and the remaining singularities can be treated by flow equations
derived from a suitable renormalization group.

In this work we analyze low energy singularities in the ground state
of charge neutral {\em fermionic} superfluids. We assume spin-singlet 
pairing with $s$-wave symmetry, so that the fermionic degrees of 
freedom are fully gapped in the superfluid phase.
The attractive Hubbard model serves as a prototypical microscopic
model for this case.
We decouple the fermionic interaction by introducing a bosonic 
pairing field via a Hubbard Stratonovich transformation, and study
the resulting coupled fermion-boson system by flow equations derived
from the functional renormalization group (fRG).

In its one-particle irreducible version,\cite{wetterich93} the fRG 
yields an exact flow equation for the effective action $\Gam^{\Lam}$, 
which interpolates smoothly between the bare action $\cS$ at the 
highest scale $\Lam_0$ and the final effective action $\Gam$ for 
$\Lam \to 0$.
Truncations of the exact flow equation have been established as
a valuable source of powerful approximation schemes in quantum 
field theory and quantum many-body physics.
\cite{berges02,kopietz10,metzner12}
Spontaneous symmetry breaking in interacting Fermi systems can 
be treated either by a fermionic flow with anomalous propagators 
and interactions,\cite{salmhofer04} or by a coupled flow involving 
the fermions and a bosonic Hubbard-Stratonovich field for the 
order parameter fluctuations.\cite{baier04}
For fermionic flows there is a relatively simple one-loop 
truncation \cite{katanin04} with the attractive feature that 
it solves mean-field models for symmetry breaking exactly, 
although the effective two-particle interaction diverges at a 
finite critical scale $\Lam_c$.
\cite{salmhofer04,gersch05,gersch06,eberlein10}
The same truncation captures also several fluctuation effects,
as has been shown in applications to the superfluid ground state 
of the attractive Hubbard model.\cite{gersch08,eberlein13}
However, the singular renormalization of longitudinal fluctuations
by the Goldstone mode is captured only in a two-loop truncation
of the fermionic flow.\cite{eberlein13a}

The treatment of order parameter fluctuations is facilitated 
by introducing a bosonic field via a Hubbard-Stratovonich 
transformation.\cite{popov87}
Fluctuation effects in the superfluid phase of attractively 
interacting Fermi systems have been studied by fRG flows with 
a Hubbard-Stratonovich field already in several works.
\cite{birse05,diehl07,krippa07,strack08,floerchinger08,bartosch09}
The interplay between longitudinal and transverse (Goldstone) order 
parameter fluctuations was addressed by Strack et al.\cite{strack08}
In particular, it was shown that the singular renormalization of
the longitudinal fluctuations known from the interacting Bose gas
is captured already in a simple one-loop truncation. 
However, the cancellations protecting the Goldstone mode were
implemented by hand, by discarding fluctuation contributions to
the transverse mass and wave function renormalization.
In the present work we complete the treatment of singular
renormalizations in a fermionic superfluid by deriving flow
equations from an improved ansatz for the effective action 
$\Gam^{\Lam}$, which captures the renormalization of longitudinal 
and transverse fluctuations in full agreement with the behavior 
of the interacting Bose gas.

This article is structured as follows.
In Sec.~II we define the bare action and introduce the 
Hubbard-Stratonovich field.
The ansatz for the effective action is formulated in Sec.~III,
and the corresponding truncated flow equations are derived in
Sec.~IV.
In Sec.~V we discuss cancellations and the asymptotic low-energy
behavior, and we present numerical results for the flow in two
dimensions.
A short conclusion in Sec.~VI closes the presentation.

%%%%%%%%%%%%%%%%%%%%%%%%%%%%%%%%%%%%%%%%%%%%%%%%%%%%%%%%%%%%%%%%%%%%%%%%%%%

\section{Bare action}

We consider a system of spin-$\frac{1}{2}$ fermions with an
attractive interaction leading to s-wave pairing. 
The precise form of the interaction is not important for the
qualitative low-energy behavior. We therefore assume a local
interaction for simplicity.
The bare action of the system is then given by
\begin{eqnarray} \label{bare_action1}
  \cS[\psi,\psib]
  & = & - \int_{k\sg}
  \psib_{k\sg} (ik_{0}-\xi_{\bk}) \, \psi_{k\sg} \nonumber \\
  && + \int_{k,k',q} U \, 
  \psib_{-k+\frac{q}{2}\down} \psib_{k+\frac{q}{2} \up}
  \psi_{k'+\frac{q}{2}\up} \psi_{-k'+\frac{q}{2}\down} \; ,
\end{eqnarray}
where $\xi_{\bk} = \eps_{\bk} - \mu$ is the single-particle energy
relative to the chemical potential, and $U < 0$ parametrizes the
attractive interaction;
$\psib_{k\sg}$ and $\psi_{k\sg}$ are fermionic (Grassmann) fields
corresponding to creation and annihiliation operators.
The variables $k = (k_0,\bk)$ and $q = (q_0,\bq)$ collect Matsubara 
energies and momenta, and $\sg$ is the spin orientation.
We use the short-hand notation
$\int_k = \int_{k_0} \int_{\bk} = 
 \int_{-\infty}^{\infty} \frac{d k_{0}}{2\pi}
 \int_{-\pi}^{\pi} \frac{d^d \bk}{(2\pi)^d} \,$
for momentum and energy integrals, 
and $\int_{k\sg}$ includes also a spin sum. 
We analyze only ground state properties, such that the energy 
variables are continuous.

In the absence of a lattice the dispersion relation is 
$\eps_{\bk} = \bk^2/2m$, and a suitable regularization of the 
action at large momenta is required.
For lattice fermions, Eq.~(\ref{bare_action1}) is the action of 
the attractive Hubbard model.\cite{micnas90}
For nearest neighbor hopping on a $d$-dimensional simple cubic 
lattice with an amlitude $-t$, the dispersion relation has the form
$\eps_{\bk} = -2t \sum_{i=1}^d \cos k_i$.

The attractive interaction causes spin-singlet pairing with s-wave
symmetry and a spontaneous breaking of the global U(1) particle
number symmetry.
We decouple the interaction in the pairing channel by introducing 
a complex bosonic Hubbard-Stratonovich field $\phi_q$ conjugate 
to the bilinear composite of fermionic fields
$\tilde\phi_{q} = U \int_k \psi_{k+\frac{q}{2}\up}
 \psi_{-k+\frac{q}{2}\down}$.
The system is then described by a functional integral over the 
fields $\psi$, $\psib$ and $\phi$, with the fermion-boson action
\begin{eqnarray} \label{bare_action2}
 \cS[\psi,\psib,\phi]
  &=& - \int_{k\sg} \psib_{k\sg} (ik_{0} - \xi_{\bk}) \, \psi_{k\sg}
  - \int_q \phi^*_q \frac{1}{U} \phi_q \nonumber \\
  && + \int_{k,q} \left( 
  \psib_{-k+\frac{q}{2} \down} \psib_{k+\frac{q}{2} \up} \,
  \phi_{q} +
  \psi_{k+\frac{q}{2} \up} \psi_{-k+\frac{q}{2}\down} \,
  \phi^*_q \right) , \hskip 5mm
\end{eqnarray}
where $\phi^*$ is the complex conjugate of $\phi$, while $\psi$
and $\psib$ are algebraically independent Grassmann variables.

%%%%%%%%%%%%%%%%%%%%%%%%%%%%%%%%%%%%%%%%%%%%%%%%%%%%%%%%%%%%%%%%%%%%%%

\section{Effective action}

Our aim is to devise a simple approximation for the fermionic and 
bosonic correlation functions that fully captures the low energy 
singularities in the superfluid ground state.
All properties of the system are encoded in the effective action 
$\Gam[\psi,\psib,\phi]$, which can be obtained by functional 
integration of the bare action in the presence of source fields 
and a subsequent Legendre transform with respect to these fields.
\cite{negele87}
Functional derivatives of $\Gam[\psi,\psib,\phi]$ with respect to 
$\psi$, $\psib$, $\phi$ (and $\phi^*$) yield the one-particle 
irreducible vertex functions, from which correlation (or Green)
functions can be easily obtained.
To analyze $\Gam[\psi,\psib,\phi]$ we use the fRG.
\cite{berges02,kopietz10,metzner12}
In that approach, one introduces a suitable infrared cutoff
to define a scale dependent effective action,
$\Gam^{\Lam}[\psi,\psib,\phi]$, which interpolates smoothly 
between the bare action $\cS[\psi,\psib,\phi]$ and the final 
effective action $\Gam[\psi,\psib,\phi]$.
The evolution of $\Gam^{\Lam}[\psi,\psib,\phi]$ as a function of 
the scale $\Lam$ is given by an exact functional flow equation.
\cite{wetterich93}

The exact effective action is a non-polynomial functional of
the fields. 
However, the low-energy behavior can be described by an 
approximate polynomial ansatz for $\Gam^{\Lam}[\psi,\psib,\phi]$, 
whose flow is given by a managable system of ordinary 
differential equations.
We keep only the leading low-energy terms which are necessary
to capture the asymptotic low-energy behavior. We make sure
that the $U(1)$ symmetry is respected by the ansatz, as is 
crucial for obtaining cancellations of unphysical divergences.
The shape of $\Gam^{\Lam}[\psi,\psib,\phi]$ changes qualitatively
at a finite critical scale $\Lam_c$, the scale of spontaneous
symmetry breaking.\cite{birse05,diehl07,krippa07,strack08}
We thus distinguish between the {\em symmetric regime} above 
$\Lam_c$ and the {\em symmetry-broken regime} below.

%%%%%%%%%%%%%%%%%%%%%%%%%%%%%%%%%%%%%%%%%%%%%%%%%%%%%%%%%%%%%%%%%%

\subsection{Symmetric regime}

In the symmetric regime, the ansatz for the scale dependent 
effective action has the form,
\begin{equation}
 \Gam^{\Lam} = \Gam_{\psib\psi} + \Gam_{\phi^*\phi} + 
 \Gam_{\phi^4} + \Gam_{\psi^2\phi^*} \; ,
\end{equation}
where the subscripts indicate the field-monomials contained in
each term.
To simplify the notation, we do not write a superscript for 
the $\Lam$-dependence of each term.

For the fermionic part, we just keep the term contained already 
in the bare action,
\begin{equation} \label{Gam_psibpsi}
 \Gam_{\psib\psi} = \int_{k\sg} 
 \psib_{k\sg} \, (\xi_{\bk} - ik_0) \, \psi_{k\sg} \; .
\end{equation}
We neglect renormalizations of the fermionic propagator since 
these are finite and do not affect the low-energy singularities 
qualitively.
We also discard fermionic interaction terms which are generated
by contributions of fourth order in the fermion-boson vertex in 
the flow. One could deal with such terms by a dynamical decoupling
of the interactions at each scale,\cite{gies02,floerchinger09}
but they also lead only to finite renormalizations.

The quadratic bosonic part is parametrized as
\begin{equation}
 \Gam_{\phi^*\phi} = \frac{1}{2} \int_q 
 \phi_q^* \, \big[m_b^2 + Z_b (q_0^2 + \om_{\bq}^2) 
 - iW q_0 \big] \, 
 \phi_q \; ,
\end{equation}
where $m_b$, $Z_b$ and $W$ are $\Lam$-dependent numbers and
$\om_{\bq}$ is a fixed function which is proportional to $|\bq|$
for small $\bq$. For a lattice system it is convenient to
choose a periodic function for $\om_{\bq}$. 
In numerical evaluations of the flow for the two-dimensional 
Hubbard model we will use
$\om_{\bq}^2 = 4t - 2t (\cos q_x + \cos q_y)$.
The bosonic mass $m_b$ is initially (in the bare action) 
$\sqrt{2/|U|}$, and vanishes at the critical scale $\Lam_c$.
The other terms are generated by fermionic pairing fluctuations,
and then renormalized also by bosonic fluctuations (see Sec.~IV).
The imaginary part proportional to $q_0$ vanishes for particle-hole
symmetric systems and was not taken into account in the ansatz used 
in Ref.~\onlinecite{strack08}.
Here we include it since this term leads to a mixing of longitudinal
and transverse fluctuations in the symmetry-broken regime, and
thus affects the low-energy singularities qualitatively.
The quadratic frequency dependence was discarded in several earlier 
works.\cite{birse05,diehl07,krippa07}
Linear and quadratic frequency terms were both taken into account in 
a previous fRG study of interacting Bose systems.\cite{floerchinger08a}
Momentum and energy dependences beyond quadratic order are 
irrelevant.

The quartic bosonic part has the form
\begin{equation}
 \Gam_{\phi^4} = \frac{1}{8} \int_{q,q',p} 
 U(p) \, \phi_{q-p}^* \phi_{q'+p}^* \phi_{q'} \phi_{q} \; .
\end{equation}
The bosonic potential is parametrized by the ansatz
\begin{equation} \label{Up}
 U(p) = u + Y \left( p_0^2 + \om_{\bp}^2 \right) \; ,
\end{equation}
where $u$ and $Y$ are scale dependent numbers.
$U(p)$ should not be confused with the Hubbard $U$ in 
Eq.~(\ref{bare_action1}).
The $u$-term corresponds to the standard local $\phi^4$ interaction.
The $Y$-term with $\om_{\bp}$ expanded to quadratic order corresponds 
to a quartic gradient term proportional to 
$\int dx \, ( \nabla |\phi(x)|^2 )^2$, where $x = (\tau,\br)$ is 
an imaginary time and (real) space coordinate.
This term is irrelevant in naive power-counting and was therefore not
included in earlier truncations of the effective action.
However, a term of this form is crucial in the symmetry-broken regime 
to implement distinct gradient terms for longitudinal and transverse
fluctuation in a $U(1)$ symmetric way, as is well known from studies 
of $O(N)$ field theories.\cite{tetradis94,strack09}
The momentum and frequency dependence of the quartic interaction
has also been taken into account in recent truncations of the effective
action for the interacting Bose gas.\cite{sinner09,dupuis09,sinner10}

The fermion-boson interaction is restricted to a 3-point vertex of
the form already present in the bare action,
\begin{equation}
 \Gam_{\psi^2\phi^*} = g \int_{k,q} 
 \big( \psib_{-k+q/2 \down} \psib_{k+q/2 \up} \phi_q +
 \psi_{k+q/2 \up} \psi_{-k+q/2 \down} \phi_q^* \big) \; .
\end{equation}
The coupling constant $g$ is equal to one in the bare action.
Within the above ansatz for the effective action it remains
unrenormalized in the symmetric regime of the flow.\cite{strack08}
Higher order fermion-boson interactions (beyond three-point) and 
momentum or energy dependences are irrelevant for the low-energy
behavior.

%%%%%%%%%%%%%%%%%%%%%%%%%%%%%%%%%%%%%%%%%%%%%%%%%%%%%%%%%%%%%%%%%%%%

\subsection{Symmetry-broken regime}

Below the critical scale $\Lam_c$ the $U(1)$ symmetry of the system
is spontaneously broken, so that anomalous terms emerge in the
effective action.
The bosonic part of the action develops a minimum at a finite
$\phi_{q=0} = \alf$, where only the modulus of $\alf$ is fixed by 
the minimization condition while the phase remains arbitrary.
Implementing this minimum in our ansatz for the effective action,
one is led to the following form for the bosonic part,
\begin{eqnarray} \label{Gamb1}
 \Gam_b &=& 
 \frac{Z_b}{2} \int dx \, |\nabla\phi(x)|^2 + 
 \frac{W}{2} \int dx \, \phi^*(x) \partial_{\tau} \phi(x)
 \nonumber \\
 &+& \frac{u}{8} \int dx \, 
 \left( |\phi(x)|^2 - |\alf|^2 \right)^2 +
 \frac{Y}{8} \int dx \, 
 \left( \nabla |\phi(x)|^2 \right)^2 ,
\end{eqnarray}
where $x = (\tau,\br)$.
Here we have written $\Gam_b$ as a functional of $\phi(x)$ because 
the structure of the quartic terms is more transparent in a
space-time (instead of momentum-frequency) representation. 
All terms in $\Gam_b$ obviously respect the global $U(1)$ symmetry.

We now fix the phase of the superfluid order parameter by choosing
$\alf$ real and positive, and we decompose $\phi(x)$ as
\begin{equation}
 \phi(x) = \alf + \sg(x) + i\pi(x) \; ,
\end{equation}
where $\sg(x)$ and $\pi(x)$ are real fields describing longitudinal
and transverse fluctuations, respectively.
In Fourier space, the decomposition reads
\begin{eqnarray}
 \phi_q &=& \alf\delta_{q0} + \sg_q + i\pi_q \; ,
 \nonumber \\
 \phi_q^* &=& \alf\delta_{q0} + \sg_{-q} - i\pi_{-q} \; ,
\end{eqnarray}
where $\sg_q^* = \sg_{-q}$ and $\pi_q^* = \pi_{-q}$ since
$\sg(x)$ and $\pi(x)$ are real.
Inserting the decomposition of $\phi$ into Eq.~(\ref{Gamb1}),
one obtains several quadratic, cubic and quartic terms,
\begin{eqnarray} \label{Gamb2}
 \Gam_b &=& 
 \Gam_{\sg^2} + \Gam_{\pi^2} + \Gam_{\sg\pi} + \Gam_{\pi\sg}
 \nonumber \\ 
 &+& \Gam_{\sg^3} + \Gam_{\sg\pi^2} +
     \Gam_{\sg^4} + \Gam_{\pi^4} + \Gam_{\sg^2\pi^2} \; .
\end{eqnarray}
The quadratic terms have the form
\begin{eqnarray}
 \Gam_{\sg^2} &=& \frac{1}{2} \int_q
 \big[ m_{\sg}^2 + Z_{\sg} (q_0^2 + \om_{\bq}^2) \big] \,
 \sg_q \sg_{-q} \; ,
 \nonumber \\
 \Gam_{\pi^2} &=& \frac{1}{2} \int_q
 Z_{\pi} (q_0^2 + \om_{\bq}^2) \, \pi_q \pi_{-q} \; ,
 \nonumber \\
 \Gam_{\sg\pi} &=& 
 - \frac{1}{2} \int_q W q_0 \, \sg_q \pi_{-q} \; ,
 \nonumber \\
 \Gam_{\pi\sg} &=& 
 \frac{1}{2} \int_q W q_0 \, \pi_q \sg_{-q} \; .
\end{eqnarray}
Note that we have returned to the momentum representation and
replaced the factors $\bq^2$ corresponding to spatial Laplace
operators by the function $\om_{\bq}^2$ introduced in Sec.~IIIA.
The term with the first order time derivative in Eq.~(\ref{Gamb1})
leads to a mixing of $\sg$ and $\pi$ fields. 
The longitudinal mass is determined by the order parameter 
$\alf$ and the quartic coupling $u$ in Eq.~(\ref{Gamb1}) as
\begin{equation} \label{rel_mualf}
 m_{\sg}^2 = u \alf^2 \; .
\end{equation}
The $Z$-factors for longitudinal and transverse fluctuations 
are related to $Z_b$ and $Y$ by
\begin{eqnarray} \label{rel_ZY}
 Z_{\sg} &=& Z_b + Y\alf^2 \; , 
 \nonumber \\
 Z_{\pi} &=& Z_b \; .
\end{eqnarray}
The cubic and quartic interaction terms read
\begin{eqnarray} \label{rel_vertex}
 \Gam_{\sg^3} &=& \frac{1}{2} \int_{q,p} 
 U(p) \, \alf \, \sg_p \sg_q \sg_{-q-p} \; , \nonumber \\
  \Gam_{\sg\pi^2} &=& \frac{1}{2} \int_{q,p} 
 U(p) \, \alf \, \sg_p \pi_q \pi_{-q-p} \; , \nonumber \\
 \Gam_{\sg^4} &=& \frac{1}{8} \int_{q,q',p}
 U(p) \, \sg_q \sg_{p-q} \sg_{q'} \sg_{-p-q'} \; , \nonumber \\
 \Gam_{\pi^4} &=& \frac{1}{8} \int_{q,q',p}
 U(p) \, \pi_q \pi_{p-q} \pi_{q'} \pi_{-p-q'} \; , \nonumber \\
 \Gam_{\sg^2\pi^2} &=& \frac{1}{4} \int_{q,q',p}
 U(p) \, \sg_q \sg_{p-q} \pi_{q'} \pi_{-p-q'} \; ,
\end{eqnarray}
where $U(p) = u + Y (p_0^2 + \om_{\bp}^2)$ as in Eq.~(\ref{Up}).
The above contributions to $\Gam_b$ are the same as for
longitudinal and transverse fields in an $O(2)$ model.
\cite{tetradis94,strack09}
The relations (\ref{rel_mualf})-(\ref{rel_vertex}) are valid
for the quartic ansatz (\ref{Gamb1}). Expanding the exact
effective action around $\alf$ would lead to additional terms
from higher (than quartic) orders.

Spontaneous symmetry breaking leads also to anomalous fermionic
contributions. The normal quadratic term $\Gam_{\psib\psi}$, 
Eq.~(\ref{Gam_psibpsi}), is supplemented by the anomalous term
\begin{equation}
 \Gam_{\psi\psi} = \int_k \left(
 \Delta \psib_{-k\down} \psib_{k\up} + 
 \Delta^* \psi_{k\up} \psi_{-k\down} \right) \; ,
\end{equation}
where $\Delta$ is the fermionic excitation gap.
Furthermore, an anomalous fermion-boson interaction of the
form $\Gam_{\psi^2\phi} = \tilde g \int_{k,q}
 (\psib_{-k+q/2\down} \psib_{k+q/2\up} \phi_{-q}^* + 
 \psi_{k+q/2\up} \psi_{-k+q/2\down} \phi_{-q} )$
is generated by the flow for $\Lam < \Lam_c$.\cite{strack08}
Decomposing $\phi$ in longitudinal and transverse fields
one obtains the fermion-boson interaction in the form
\begin{eqnarray}
 \Gam_{\psi^2\sg} &=& g_{\sg} \int_{k,q} \left(
 \psib_{-k+q/2\down} \psib_{k+q/2\up} \sg_q + 
 \psi_{k+q/2\up} \psi_{-k+q/2\down} \sg_{-q} \right) \; , \\
 \Gam_{\psi^2\pi} &=& ig_{\pi} \int_{k,q} \left(
 \psib_{-k+q/2\down} \psib_{k+q/2\up} \pi_q - 
 \psi_{k+q/2\up} \psi_{-k+q/2\down} \pi_{-q} \right) \; ,
\end{eqnarray}
where $g_{\sg} = g + \tilde g$ and $g_{\pi} = g - \tilde g$.
A Ward identity derived from the $U(1)$ symmetry yields a
relation between the fermionic gap and the bosonic order
parameter $\alf$, namely
\begin{equation} \label{WI}
 \Delta = g_{\pi} \alf \; .
\end{equation}
A derivation is given in Appendix A. A similar relation for
a truncation with $\tilde g = 0$ was derived previously in 
Ref.~\onlinecite{bartosch09}. For $\tilde g = 0$, there is only
one Yukawa coupling, $g_{\pi} = g_{\sg} = g$, and the Ward 
identity can be written as $\Delta = g \alf$.

In summary, our ansatz for the effective action in the 
symmetry-broken regime has the form
\begin{equation}
 \Gam^{\Lam} = \Gam_b + \Gam_{\psib\psi} + \Gam_{\psi\psi} +
 \Gam_{\psi^2\sg} + \Gam_{\psi^2\pi} \; ,
\end{equation}
where $\Gam_b$ is a sum of purely bosonic terms as listed in
Eq.~(\ref{Gamb2}).
An ansatz with distinct renormalizations for longitudinal and
transverse fluctuation fields was formulated and evaluated 
already in Ref.~\onlinecite{strack08}.
Here we have included two additional terms. Most importantly,
the quartic gradient term ($Y$-term) is crucial for capturing
the very different behavior of $Z_{\sg}$ and $Z_{\pi}$ while
conserving the $U(1)$ symmetry.\cite{tetradis94,strack09}
Eq.~(\ref{rel_ZY}) implies that setting $Y=0$ forces 
$Z_{\sg} = Z_{\pi}$ within a $U(1)$ symmetric ansatz. 
Implementing $Z_{\sg} \neq Z_{\pi}$ at the expense of the $U(1)$ 
symmetry as in Ref.~\onlinecite{strack08} spoils important 
cancellations of singularities, so that the Goldstone mode 
has to be protected by hand.
Second, the term linear in $q_0$ in the bosonic sector
($W$-term) leads to a mixing of $\sg$ and $\pi$ fields in
the symmetry-broken regime. This term is important for 
capturing the generic low-energy behavior of the superfluid.
In particular, it determines the condensate compressibility.
\cite{pistolesi04}

%%%%%%%%%%%%%%%%%%%%%%%%%%%%%%%%%%%%%%%%%%%%%%%%%%%%%%%%%%%%%%%%%%%%

\section{Flow equations}

Inserting the ansatz for the effective action $\Gam^{\Lam}$
into the exact flow equation \cite{berges02,kopietz10,metzner12}
and comparing coefficients, one obtains flow equations for the 
scale dependent parameters.
The scale dependence is generated by regulator functions which
are added to the inverse propagators.
For the fermions we choose the function
\begin{equation}
 R_f(k) = R_f(k_0) =
 [i\Lam \sgn(k_0) - ik_0] \, \Theta(\Lam - |k_0|) \; ,
\end{equation}
and for the bosons
\begin{equation}
 R_b(q) = R_b(q_0) =
 Z_b (\Lam^2 - q_0^2) \, \Theta(\Lam - |q_0|) \; .
\end{equation}
Regulator functions of this form lead to relatively simple
integrals.\cite{litim01} The $U(1)$ symmetry is not spoiled
by these regulators.
The scale derivatives of the regulators are
\begin{equation}
 \frac{d}{d\Lam} R_f(k) = 
 i \sgn(k_0) \Theta(\Lam - |k_0|) \; ,
\end{equation}
and
\begin{equation}
 \frac{d}{d\Lam} R_b(q) = 
 \left[ 2 Z_b \Lam + \frac{dZ_b}{d\Lam}(\Lam^2 - q_0^2) \right]
 \Theta(\Lam - |q_0|) \; .
\end{equation}
The scale derivative of $Z_b$ turns out to be quite large
at the initial stage of the flow. 
Hence, we take its contribution to the scale derivative of
the regulator into account.
We now present the flow equations in the symmetric and 
symmetry-broken regime.

%%%%%%%%%%%%%%%%%%%%%%%%%%%%%%%%%%%%%%%%%%%%%%%%%%%%%%%%%%%%

\subsection{Symmetric regime}

The right hand sides of the flow equations are loop integrals 
over products of fermionic and bosonic propagators. 
The propagators are the inverse of the quadratic kernels 
of the effective action (including the regulator functions).
In the symmetric regime (for $\Lam > \Lam_c$), the fermion 
propagator has the form
\begin{equation}
 G_f(k) = - \bra \psi_{k\sg} \psib_{k\sg} \ket =
 \frac{1}{ik_0 - \xi_{\bk} + R_f(k)} \; ,
\end{equation}
and the boson propagator reads \cite{fn1}
\begin{equation}
 G_b(q) = \frac{1}{2} \, \bra \phi_q \phi_{q}^* \ket =
 \frac{1}{m_b^2 + Z_b(q_0^2 + \om_{\bq}^2) - iW q_0 + R_b(q)} 
 \; .
\end{equation}
The regulator functions replace $k_0$ by $\sgn(k_0) \, \Lam$
in the fermion propagator and $q_0^2$ by $\Lam^2$ in the
boson propagator, if $|k_0| < \Lam$ and $|q_0| < \Lam$, 
respectively.
One factor in the propagator product is subject to a scale
derivative acting on the regulator function, leading to 
the single-scale propagators
\begin{equation} 
 G'_{f/b}(k) = D_{\Lam} G_{f/b}(k) 
 = - G_{f/b}^2(k) \, \partial_{\Lam} R_{f/b}(k)
\end{equation}
for fermions and bosons, respectively. The differential
operator $D_{\Lam}$ acts only on the regulator function,
not on other $\Lam$-dependences of the propagators.

The flow equations for the bosonic mass, the bosonic 
$Z$-factor, and the coefficient $W$ are obtained from the
flow of the bosonic self-energy
$\Sg_b(p) = [G_b(p)]^{-1} - [G_b^0(p)]^{-1}$, with
$G_b^0(p) = [m_b^0 + R_b(p)]^{-1}$.
The self-energy obeys the flow equation
\begin{eqnarray}
 \frac{d}{d\Lam} \Sg_b(p) &=&
 - 2 g^2 \int_k D_{\Lam} \left [ G_f(k) G_f(p-k) \right]
 \nonumber \\
 &+& \int_{q} U(0) \, G'_b(q) + U(p-q) \, G'_b(q) \; .
\end{eqnarray}
The first term is a fermionic one-loop contribution, while the
second and third term is the bosonic Hartree and Fock term,
respectively.
The corresponding Feynman diagrams are shown in Fig.~1, where
the Hartree and Fock terms are represented by a single diagram
with a symmetrized vertex.
\begin{figure}[tb]
\centerline{\includegraphics[width=5cm]{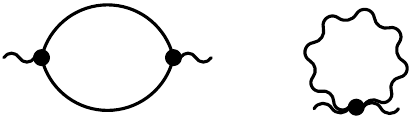}}
\caption{Feynman diagrams representing the contributions to
 the flow of the bosonic self-energy in the symmetric regime.
 Solid and wiggly lines represent fermion and boson propagators,
 respectively. Dots correspond to vertices. The internal 
 structure (momentum and frequency dependence) of the bosonic 
 interaction vertex is not displayed.}
\end{figure}
The flow of the mass is obtained from the flow of $\Sg_b(0)$ as
\begin{eqnarray}
 \frac{d}{d\Lam} m_b^2 &=& 
 \frac{d}{d\Lam} \Sg_b(0) =
 - 4 g^2 \int_k G'_f(k) G_f(-k) \nonumber \\
 &+& \int_{q} [u + U(q)] \, G'_b(q) \; .
\end{eqnarray}
The fermionic term reduces the mass upon lowering $\Lam$,
while the bosonic term partially compensates this reduction.
The flow for the bosonic $Z$-factor is extracted by applying
a second order momentum derivative to the bosonic self-energy,
that is,
$\frac{d}{d\Lam} Z_b = \frac{1}{2} \partial_{p_x}^2 \left.
 \frac{d}{d\Lam} \Sg_b(p) \right|_{p=0}$.\cite{fn2}
This yields
\begin{eqnarray}
 \frac{d}{d\Lam} Z_b &=& 
 - 2 g^2 \partial_{p_x}^2 \int_k 
 \left. G'_f(k) G_f(p-k) \, \right|_{q=0} \nonumber \\
 &+& \frac{Y}{2} \, \partial_{p_x}^2 \int_{q} 
 \left. \om_{\bp-\bq}^2 \, G'_b(q) \, \right|_{p=0} \; .
\end{eqnarray}
Note that the second momentum derivative at $p=0$ does not
depend on the direction, even for lattice systems.
The flow of the parameter $W$ is obtained from a frequency 
derivative, 
$\frac{d}{d\Lam} W = i \partial_{p_0} \left.
 \frac{d}{d\Lam} \Sg_b(p) \right|_{p=0} \,$,
yielding
\begin{equation}
 \frac{d}{d\Lam} W = 
 - 4i g^2 \partial_{p_0} \int_k 
 \left. G'_f(k) G_f(p-k) \, \right|_{p=0} 
 - 2iY \int_{q} q_0 \, G'_b(q) \; .
\end{equation}

The flow of the quartic couplings $u$ and $Y$ is determined
by the flow of the bosonic four-point vertex.
The Feynman diagrams corresponding to the fermionic and bosonic 
one-loop contributions are shown in Fig.~2.
\begin{figure}[tb]
\centerline{\includegraphics[width=5cm]{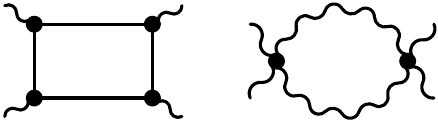}}
\caption{Feynman diagrams representing the contributions to
 the bosonic interaction vertex in the symmetric regime.}
\end{figure}
Setting all ingoing and outgoing momenta to zero, one
obtains the flow equation for $u$,
\begin{eqnarray} \label{floweq_u}
 \frac{d}{d\Lam} u &=& 4 g^4 \int_k D_{\Lam} 
 \left[ G_f^2(k) G_f^2(-k) \right] \nonumber \\
 &-& \int_q \left[ U(q) \right]^2 \, 
 D_{\Lam} \left[ G_b(q) G_b(-q) \right] \nonumber \\
 &-& \int_q \left[ U(0) + U(q) \right]^2 \, 
 D_{\Lam} \left[ G_b(q) G_b(q) \right] \; .
\end{eqnarray}
The flow equation for $Y$ is obtained by setting ingoing 
and outgoing momenta equal to $\pm p/2$ such that the 
total momentum is zero.
Applying a second order derivative with respect to $p_x$ 
to the flow equation for the (symmetrized) vertex yields
\begin{eqnarray} \label{floweq_Y}
 \frac{d}{d\Lam} Y &=& 4 g^4 \partial_{p_x}^2 \int_k
 \left. D_{\Lam} \left[ G_f^2(k) G_f(-k+p/2) G_f(-k-p/2) 
 \right] \, \right|_{p=0}
 \nonumber \\
 &-&  \frac{1}{4} \, \partial_{p_x}^2 \int_q 
 \left[ U(q+p/2) + U(q-p/2) \right]^2 \, 
 \nonumber \\ &&\times
 \left. D_{\Lam} \left[ G_b(q) G_b(-q) \right] \, \right|_{p=0}
 \nonumber \\
 &-&  \frac{1}{2} \, \partial_{p_x}^2 \int_q 
 \left[ U(0) + U(q+p/2) \right] \, 
 \left[ U(0) + U(q-p/2) \right] \,
 \nonumber \\ &&\times 
 \left. D_{\Lam} \left[ G_b(q) G_b(q) \right] \, \right|_{p=0} 
 \nonumber \\
 &-&  \frac{1}{2} \, \partial_{p_x}^2 \int_q 
 \left[ U(p) + U(q) \right]^2
 \nonumber \\ &&\times 
 \left. D_{\Lam} \left[ G_b(q+p/2) G_b(q-p/2) \right] \, 
 \right|_{p=0} \; .
\end{eqnarray}
The first contribution in Eqs.~(\ref{floweq_u}) and 
(\ref{floweq_Y}) is generated by fermions, while the 
remaining terms are due to bosonic fluctuations.

%%%%%%%%%%%%%%%%%%%%%%%%%%%%%%%%%%%%%%%%%%%%%%%%%%%%%%%%%%%

\subsection{Symmetry-broken regime}

Spontaneous symmetry breaking leads to anomalous terms in
the effective action and the form of the propagators changes
accordingly.
Inverting the quadratic fermionic terms yields the normal
and anomalous fermion propagators
\begin{eqnarray}
 G_f(k) &=& - \bra \psi_{k\sg} \psib_{k\sg} \ket = 
 \frac{-ik_0 - R_f(k_0) - \xi_{\bk}}
 {|ik_0 + R_f(k_0)|^2 + E_{\bk}^2} \; , \\
 F_f(k) &=& - \bra \psi_{k\up} \psi_{-k\down} \ket =
 \frac{\Delta}{|ik_0 + R_f(k_0)|^2 + E_{\bk}^2} \; ,
\end{eqnarray}
where $E_{\bk} = \sqrt{\Delta^2 + \xi_{\bk}^2}$ is the
fermionic excitation energy in the superfluid phase.
The corresponding single-scale propagators are given by
$G'_f(k) = D_{\Lam} G_f(k)$ and $F'_f(k) = D_{\Lam} F_f(k)$.
Note the relation
\begin{equation} \label{rel_GF}
 |G_f(k)|^2 + [F_f(k)]^2 = 
 \frac{1}{\Delta} F_f(k) \; .
\end{equation}

Inverting the quadratic bosonic terms yields the 
propagators for the $\sg$ and $\pi$ fields
\begin{eqnarray}
 G_{\sg\sg}(q) &=& \bra \sg_q \sg_{-q} \ket =
 \frac{\gam_{\pi\pi}(q)}
 {\gam_{\sg\sg}(q) \, \gam_{\pi\pi}(q) + W^2 q_0^2} \; , \\
 G_{\pi\pi}(q) &=& \bra \pi_q \pi_{-q} \ket =
 \frac{\gam_{\sg\sg}(q)}
 {\gam_{\sg\sg}(q) \, \gam_{\pi\pi}(q) + W^2 q_0^2} \; , \\
 G_{\pi\sg}(q) &=& \bra \pi_q \sg_{-q} \ket =
 \frac{W q_0}
 {\gam_{\sg\sg}(q) \, \gam_{\pi\pi}(q) + W^2 q_0^2} \; ,
\end{eqnarray}
and $G_{\sg\pi}(q) = - G_{\pi\sg}(q)$, where
$\gam_{\sg\sg}(q) = 
 m_{\sg}^2 + Z_{\sg} (q_0^2 + \om_{\bq}^2) + R_b(q)$ and 
$\gam_{\pi\pi}(q) = Z_{\pi} (q_0^2 + \om_{\bq}^2) + R_b(q)$.
The corresponding single-scale propagators are
$G'_{\sg\sg}(q) = D_{\Lam} G_{\sg\sg}(q)$ etc.
The relation
\begin{equation} \label{rel_gamU}
 \gam_{\sg\sg}(q) - \gam_{\pi\pi}(q) = U(q) \, \alf^2
\end{equation}
following from Eqs.~(\ref{rel_mualf}) and (\ref{rel_ZY})
will be useful below.

\begin{figure}[tb]
\centerline{\includegraphics[width=4cm]{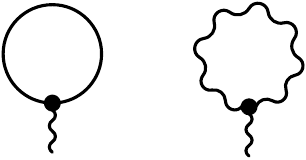}}
\caption{Feynman diagrams representing the fermionic and
 bosonic one-loop contributions to the bosonic one-point 
 vertex in the symmetry-broken regime. Here and in the
 following figures, fermionic propagators (solid lines) may
 be normal or anomalous, and bosonic propagators (wiggly 
 lines) may be $G_{\sg\sg}$, $G_{\pi\pi}$, or mixed.}
\end{figure}
The order parameter $\alf$ is determined from the 
condition that the one-point $\sg$-vertex $\gam_{\sg}$ 
has to vanish if $\alf$ is a minimum of the effective
action.
The flow of $\gam_{\sg}$ is given by
\begin{eqnarray}
 \frac{d \gam_{\sg}}{d\Lam} &=& 
 m_{\sg}^2 \frac{d\alf}{d\Lam} - 2g_{\sg} \int_k F'_f(k) 
 \nonumber \\
 &+& \frac{\alf}{2} \int_q 
 \left[ u + 2U(q) \right] \, G'_{\sg\sg}(q) 
 + \frac{\alf}{2} \int_q u \, G'_{\pi\pi}(q) \, .
\end{eqnarray}
The first term on the right hand side is generated by the
scale dependence of the point around which the effective
action is expanded (in powers of the fields).\cite{strack08}
The Feynman diagrams representing the other contributions
are shown in Fig.~3.
The condition $\frac{d}{d\Lam} \gam_{\sg} = 0$ yields the flow 
equation for $\alf$,
\begin{eqnarray} \label{floweq_alf}
 \frac{d\alf}{d\Lam} &=& 
 2 \frac{g_{\sg}}{m_{\sg}^2} \int_k F'_f(k) 
 \nonumber \\
 &-& \frac{\alf}{2m_{\sg}^2} \int_q 
 \left\{ \left[ u + 2U(q) \right] \, G'_{\sg\sg}(q) 
 + u \, G'_{\pi\pi}(q) \right\}  \, .
\end{eqnarray}

We now turn to quantities which can be derived from the 
$\sg$- and $\pi$-field self-energies. The one-loop 
contributions to the flow of these bosonic self-energies 
are represented diagrammatically in Fig.~4. 
\begin{figure}[tb]
\vskip 5mm
\centerline{\includegraphics[width=5cm]{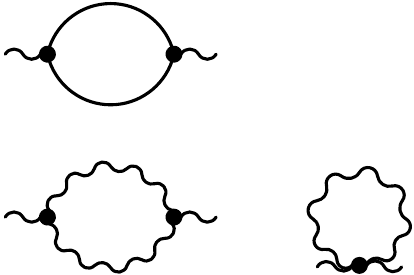}}
\caption{Feynman diagrams representing contributions to 
 the bosonic self-energies in the symmetry-broken regime.}
\end{figure}
The flow of $m_{\sg}^2$ and $Z_{\sg}$ can be extracted from 
the flow of the $\sg$-field self-energy.
The latter obeys the flow equation
\begin{widetext}
\begin{eqnarray} \label{floweq_Sg_sg}
 \frac{d}{d\Lam} \Sg_{\sg\sg}(p) &=& 
 \left[ U(0) + 2U(p) \right] \alf \frac{d\alf}{d\Lam}
 - g_{\sg}^2 \int_k \left\{ D_{\Lam} 
 \left[ G_f(k) G_f(p-k) - F_f(k) F_f(p-k) \right] 
 + (p \mapsto -p) \right\}
 \nonumber \\
 &+& \frac{1}{2} \int_q 
 \left\{ \left[ U(0) + 2U(p+q) \right] \, G'_{\sg\sg}(q) 
 + u \, G'_{\pi\pi}(q) \right\} 
 - \frac{1}{2} \int_q 
 \left[ U(p) + U(q) + U(p+q) \right]^2 \alf^2 \,
 D_{\Lam} \left[ G_{\sg\sg}(q) G_{\sg\sg}(p+q) \right]
 \nonumber \\
 &-& \frac{1}{2} \int_q 
 \left[ U(p) \right]^2 \alf^2 \,
 D_{\Lam} \left[ G_{\pi\pi}(q) G_{\pi\pi}(p+q) \right]
 - \int_q
 \left[ U(p) + U(q) + U(p+q) \right] U(p) \, \alf^2 \,
 D_{\Lam} \left[ G_{\sg\pi}(q) G_{\pi\sg}(p+q) \right] \; .
\end{eqnarray}
\end{widetext}
The first term on the right hand side is generated by the
scale dependence of $\alf$.
The flow of the mass is obtained from $\Sg_{\sg\sg}(0)$ as
\begin{eqnarray} \label{floweq_m_sg}
 \frac{d}{d\Lam} m_{\sg}^2 &=& 
 \frac{d}{d\Lam} \Sg_{\sg\sg}(0) 
 \nonumber \\
 &=& 3 u \alf \frac{d\alf}{d\Lam} 
 - 2 g_{\sg}^2 \int_k D_{\Lam} 
 \left[ |G_f(k)|^2 - F_f^2(k) \right] 
 \nonumber \\
 &+& \frac{1}{2} \int_q 
 \left\{ \left[ u + 2U(q) \right] \, G'_{\sg\sg}(q) 
 + u \, G'_{\pi\pi}(q) \right\} 
 \nonumber \\
 &-& \frac{\alf^2}{2} \int_q D_{\Lam}
 \left\{ \left[ u + 2U(q) \right]^2 \, G_{\sg\sg}^2(q) 
 + u^2 \, G_{\pi\pi}^2(q) \right\}
 \nonumber \\
 &+& \alf^2 \int_q 
 \left[ u + 2U(q) \right] u \,
 D_{\Lam} \left[ G_{\sg\pi}(q) \right]^2  \; .
\end{eqnarray}
The flow of $Z_{\sg}$ can be obtained from a second momentum
derivative of the self-energy at $p=0$, that is,
$\frac{d}{d\Lam} Z_{\sg} = \frac{1}{2} \left. \partial_{p_x}^2 
 \frac{d}{d\Lam} \Sg_{\sg\sg}(p) \right|_{p=0} \,$.
The flow equation for the $\pi$-field self-energy reads
\begin{eqnarray} \label{floweq_Sg_pi}
 \frac{d}{d\Lam} \Sg_{\pi\pi}(p) &=& 
 u \alf \frac{d\alf}{d\Lam}
\nonumber \\
 &-& g_{\pi}^2 \int_k \Big\{ D_{\Lam} 
 \left[ G_f(k) G_f(p-k) + F_f(k) F_f(p-k) \right]
\nonumber \\
 && + \, (p \mapsto -p) \Big\} 
 \nonumber \\
 &+& \frac{1}{2} \int_q 
 \left\{ \left[ U(0) + 2U(p+q) \right] \, G'_{\pi\pi}(q) 
 + u \, G'_{\sg\sg}(q) \right\} 
 \nonumber \\
 &-& \int_q 
 \left[ U(q) \right]^2 \alf^2 \,
 D_{\Lam} \left[ G_{\sg\sg}(q) G_{\pi\pi}(p+q) \right]
 \nonumber \\
 &-& \int_q U(q) U(p+q) \, \alf^2 \,
 D_{\Lam} \left[ G_{\sg\pi}(q) G_{\sg\pi}(p+q) \right] \; .
\nonumber \\
\end{eqnarray}
The flow of $Z_{\pi}$ can be extracted by applying a second 
order momentum derivative, that is,
$\frac{d}{d\Lam} Z_{\pi} = \frac{1}{2} \left. \partial_{p_x}^2 
 \frac{d}{d\Lam} \Sg_{\pi\pi}(p) \right|_{p=0} \,$.
The flow of $W$ in the symmetry broken regime can be obtained
from a frequency derivative of the mixed $\sg\pi$ self-energy,
that is, $\frac{d}{d\Lam} W = \left. \partial_{p_0} 
 \frac{d}{d\Lam} \Sg_{\sg\pi}(p) \right|_{p=0} \,$.
The flow of $\Sg_{\sg\pi}$ is determined by the flow equation
\begin{eqnarray} \label{floweq_Sg_sgpi}
 \frac{d}{d\Lam} \Sg_{\sg\pi}(p) &=& 
 i g_{\sg} g_{\pi} \! \int_k D_{\Lam} 
 \left[ G_f(k) G_f(-p-k) - G_f(k) G_f(p-k) \right] 
 \nonumber \\
 &-& \int_q U(p+q) \, G'_{\sg\pi}(q)
 \nonumber \\
 &-& \int_q 
 U(q) \left[ U(p) + U(q) + U(p+q) \right] \alf^2 \,
 \nonumber \\ &&\times \,
 D_{\Lam} \left[ G_{\sg\sg}(q) G_{\sg\pi}(p+q) \right]
 \nonumber \\
 &-& \int_q U(p) U(q) \, \alf^2 \,
 D_{\Lam} \left[ G_{\sg\pi}(q) G_{\pi\pi}(p+q) \right] \; .
\end{eqnarray}
Note that all contributions vanish at $p=0$, so that no
$\sg$-$\pi$-mixing mass term is generated.
The bosonic contributions to the flow of the bosonic 
self-energy in Eqs.~(\ref{floweq_Sg_sg}), (\ref{floweq_Sg_pi}) 
and (\ref{floweq_Sg_sgpi}) are equivalent to those derived
previously for the interacting Bose gas by Sinner et al.
\cite{sinner10}

In the fermionic sector, the gap is the only flow parameter
in our truncation.
Its flow is obtained from the flow of the anomalous fermionic
self-energy at zero frequency and a momentum $\bk_F$ on the 
Fermi surface. This yields the flow equation
\begin{eqnarray} \label{floweq_Delta}
 \frac{d}{d\Lam} \Delta &=&
 g_{\sg} \frac{d\alf}{d\Lam}  
 \nonumber \\
 &-& \int_q D_{\Lam} \left\{ F_f(k_F-q) \left[ 
 g_{\sg}^2 G_{\sg\sg}(q) - g_{\pi}^2 G_{\pi\pi}(q) \right] 
 \right\} \, ,
 \nonumber \\
\end{eqnarray}
where $k_F = (0,\bk_F)$.
The second (fluctuation) term can be represented by the 
Feynman diagram in Fig.~5.
\begin{figure}[tb]
\centerline{\includegraphics[width=3cm]{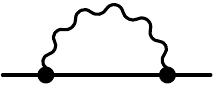}}
\caption{Feynman diagram representing the fluctuation
 contribution to the flow of the gap.}
\end{figure}
Contributions involving $\sg$-$\pi$ mixing cancel.
We recall that $D_{\Lam}$ acts only on the regulator 
functions in the propagators.
For isotropic systems, the right hand side of 
Eq.~(\ref{floweq_Delta}) does not depend on the choice of
$\bk_F$. For lattice systems the dependence is generically
very weak, since the bosonic fluctuation contribution is 
dominated by small momenta $\bq$, and 
$F_f(k_F) = \Delta/(\Delta^2 + \Lam^2)$ is independent of $\bk_F$.
Considering the Ward identity Eq.~(\ref{WI}), one may expect
the coupling $g_{\pi}$ instead of $g_{\sg}$ in the contribution
proportional to $d\alf/d\Lam$ in Eq.~(\ref{floweq_Delta}).
Indeed, we will see below that the Ward identity is fulfilled
within our truncation of the effective action only if one sets 
$g_{\sg} = g_{\pi}$.

The set of flow equations is completed by the flow of the
couplings $g_{\sg}$ and $g_{\pi}$ parametrizing the 
fermion-boson interaction. 
The flow of $g_{\pi}$ is given by
\begin{eqnarray} \label{floweq_g_pi1}
 \frac{d}{d\Lam} g_{\pi} &=&
 g_{\pi} \int_q D_{\Lam} \Big\{
 \left[ F_f^2(k_F-q) + |G_f(k_F-q)|^2 \right] 
 \nonumber \\ && \times \,
 \left[ g_{\sg}^2 G_{\sg\sg}(q) - g_{\pi}^2 G_{\pi\pi}(q)
 \right] \Big\} \nonumber \\
 &+& 2 g_{\sg} g_{\pi} \int_q U(q) \, \alf \, D_{\Lam}
 \Big\{ F(k_F - q) 
 \nonumber \\ && \times \,
 \left[ 
 G_{\sg\sg}(q) G_{\pi\pi}(q) + G_{\sg\pi}(q) G_{\sg\pi}(q) 
 \right] \Big\} \; .
\end{eqnarray}
The corresponding Feynman diagrams are shown in Fig.~6.
\begin{figure}[tb]
\centerline{\includegraphics[width=6cm]{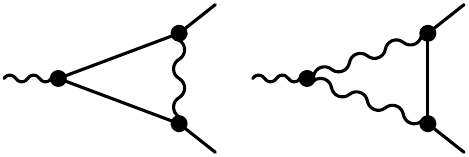}}
\caption{Feynman diagrams representing the flow of the
 fermion-boson interaction vertex.}
\end{figure}
Using the relations (\ref{rel_GF}) and (\ref{rel_gamU}),
the flow equation can be simplified to
\begin{eqnarray} \label{floweq_g_pi2}
 \frac{d}{d\Lam} g_{\pi} &=&
 \frac{g_{\pi}}{\Delta} \int_q D_{\Lam} 
 \left\{ F_f(k_F - q)
 \left[ g_{\sg}^2 G_{\sg\sg}(q) - g_{\pi}^2 G_{\pi\pi}(q)
 \right] \right\} \nonumber \\
 &-& 2 \frac{g_{\sg} g_{\pi}}{\alf} \int_q D_{\Lam}
 \left\{ F_f(k_F - q) \left[ 
 G_{\sg\sg}(q) - G_{\pi\pi}(q) \right] \right\} \; .
 \nonumber \\
\end{eqnarray}
The flow equation for $g_{\sg}$ will not be used for 
reasons that become clear below.

Our ansatz is an extension of the truncation used by Strack
et al.,\cite{strack08} where the $W$-term was absent and the
bosonic potential $U(p)$ was constant in momentum space 
(that is, $Y=0$).
Setting $W = Y = 0$, the flow equations presented above
reduce essentially to the previously derived flow equations,
with three corrections and extensions. 
First, we have obtained an extra factor $\frac{1}{2}$ in front
of the two-boson contributions to the $\sg$-field self-energy,
which determines the flow of $m_{\sg}^2$ and $Z_{\sg}$.
Second, the two-boson contribution to the flow of the 
fermion-boson vertex was overlooked in Ref.~\onlinecite{strack08}.
Third, the bosonic fluctuation contributions to $Z_{\pi}$
were discarded. 
The latter are finite and therefore not crucial qualitatively,
but they are finite only as a consequence of rather subtle 
cancellations, as we shall see below.

%%%%%%%%%%%%%%%%%%%%%%%%%%%%%%%%%%%%%%%%%%%%%%%%%%%%%%%%%%%%

\subsection{Goldstone theorem and Ward identity
 for gap}

The Goldstone theorem implies that transverse order parameter
fluctuations are massless, that is, there is no $\pi$-field mass 
$m_{\pi}$. Such a mass is indeed absent in our $U(1)$-symmetric
ansatz for the effective action. However, inserting the ansatz
into the flow equations, various contributions to $m_{\pi}^2 =
\Sg_{\pi\pi}(0)$ are being generated and it is not obvious 
that these contributions cancel.
Evaluating the flow equation (\ref{floweq_Sg_pi}) for 
$\Sg_{\pi\pi}(p)$ at $p=0$ yields
\begin{eqnarray}
 \frac{d}{d\Lam} m_{\pi}^2 &=& 
 u \alf \frac{d\alf}{d\Lam}
 - 2 g_{\pi}^2 \int_k D_{\Lam} 
 \left[ |G_f(k)|^2 + F_f^2(k) \right] 
 \nonumber \\
 &+& \frac{1}{2} \int_q 
 \left\{ \left[ u + 2U(q) \right] \, G'_{\pi\pi}(q) 
 + u \, G'_{\sg\sg}(q) \right\} 
 \nonumber \\
 &-& \!\! \int_q \left[ U(q) \right]^2 \alf^2 
 D_{\Lam} \left[ G_{\sg\sg}(q) G_{\pi\pi}(q) +
 G_{\sg\pi}^2(q) \right] . \hskip 5mm
\end{eqnarray}
Using Eqs.~(\ref{rel_GF}) and (\ref{rel_gamU}), the right
hand side can be simplified to
\begin{eqnarray}
 \frac{d}{d\Lam} m_{\pi}^2 &=& 
 u \alf \frac{d\alf}{d\Lam}
 - 2 \frac{g_{\pi}^2}{\Delta} \int_k F'_f(k) 
 \nonumber \\
 &+& \frac{1}{2} \int_q 
 \left\{ \left[ u + 2U(q) \right] \, G'_{\sg\sg}(q) 
 + u \, G'_{\pi\pi}(q) \right\} \; .
\end{eqnarray}
Inserting the flow equation for $\alf$, Eq.~(\ref{floweq_alf}),
the bosonic fluctuation contributions cancel such that
\begin{equation}
 \frac{d}{d\Lam} m_{\pi}^2 = 
 2 \left( \frac{g_{\sg}}{\alf} - \frac{g_{\pi}^2}{\Delta} \right) 
 \int_k F'_f(k) \; .
\end{equation}
Using the Ward identity $\Delta = g_{\pi} \alf$, the prefactor in front 
of the integral can be written as $\frac{2}{\alf}(g_{\sg} - g_{\pi})$.
Hence, the contributions to the flow of $m_{\pi}^2$ vanish if (and
only if) $g_{\sg} = g_{\pi}$.
Within our truncation of the effective action, we are thus constrained 
to set $g_{\sg} = g_{\pi}$ in order to respect the Goldstone theorem.
%No problem in Bose sector!

Using the condition $g_{\sg} = g_{\pi}$ and the Ward identity
$\Delta = g_{\pi} \alf$, the two contributions to the flow of $g_{\pi}$
in Eq.~(\ref{floweq_g_pi2}) can be combined to
\begin{equation} \label{floweq_g_pi3}
 \frac{d}{d\Lam} g_{\pi} =
 - \frac{g_{\pi}^2}{\alf} \int_q D_{\Lam} 
 \left\{ F_f(k_F - q) \left[ G_{\sg\sg}(q) - G_{\pi\pi}(q)
 \right] \right\} \; .
\end{equation}
Comparing this equation with the flow equation (\ref{floweq_Delta}),
one can see that the flow equations for $g_{\pi}$ and $\Delta$ are 
indeed consistent with the Ward identity $\Delta = g_{\pi} \alf$, if 
one chooses $g_{\sg} = g_{\pi}$.
The Ward identity would not be respected by the flow if one computed
$g_{\sg}$ and $g_{\pi}$ from their distinct flow equations, or if both
were set equal and computed from the flow equation for $g_{\sg}$.

In summary, the flow equations are consistent with the Goldstone
theorem ($m_{\pi} = 0$) and the Ward identity $\Delta = g_{\pi} \alf$,
if one sets $g_{\sg} = g_{\pi}$, and computes the flow of this unified
coupling from the flow equation for $g_{\pi}$.
Since $g_{\pi}$ deviates only mildly from its initial value 
$g_{\pi}^0 = 1$, one might also simplify the truncation further by 
discarding the flow of the fermion-boson vertex completely, such that 
$g_{\sg} = g_{\pi} = 1$ and $\Delta = \alf$.
To implement $g_{\sg} \neq g_{\pi}$ consistently, one would have to
extend the ansatz for the effective action, as discussed briefly
in the Conclusions.

%%%%%%%%%%%%%%%%%%%%%%%%%%%%%%%%%%%%%%%%%%%%%%%%%%%%%%%%%%%%%%%%%%%%%%%

\section{Flow}

We now discuss the behavior of the flow as obtained from the flow
equations derived in the preceding section.
We first analyze the low energy behavior in the limit $\Lam \to 0$
and then present numerical results for the flow on all scales,
using the two-dimensional Hubbard model as a prototypical example.

%%%%%%%%%%%%%%%%%%%%%%%%%%%%%%%%%%%%%%%%%%%%%%%%%%%%%%%%%%%%%%%%%%%%%%%

\subsection{Low energy behavior}

The low energy behavior in the limit $\Lam \to 0$ is independent of
model details and can be analyzed quite generally.
For $\Lam < \Lam_c$ the fermionic propagator is regularized by the 
energy gap $\Delta$. Infrared singularities arise therefore solely
from the bosonic propagators, where the most singular one is the
propagator for transverse order parameter fluctuations $G_{\pi\pi}$.

It is well-known that the Goldstone mode leads to a singular 
renormalization of longitudinal fluctuations.
\cite{nepomnyashchy92,pistolesi04}
For $\Lam \to 0$, the flow equation (\ref{floweq_m_sg}) for 
$m_{\sg}^2$ is dominated by the term involving a product of two
$\pi$-propagators, such that
\begin{equation} \label{asymp_m_sg}
 \frac{d}{d\Lam} m_{\sg}^2 \sim
 - \alf^2 u^2 \int_q G_{\pi\pi}(q) \, G'_{\pi\pi}(q) \; .
\end{equation}
The integral diverges as $\Lam^{d-4}$ for $\Lam \to 0$, where
$d$ is the spatial dimensionality.
Note that $G'_{\pi\pi}(q)$ vanishes for $|q_0| > \Lam$.
Using the relation $m_{\sg}^2 = u \alf^2$ one thus finds that
$m_{\sg}^2$ and $u$ both vanish as $\Lam^{3-d}$ for $d < 3$, and
logarithmically in three dimensions,\cite{strack08} in agreement
with the behavior derived by other methods.
\cite{nepomnyashchy92,pistolesi04}

The flow of $Z_{\sg}$ is also dominated by the two-$\pi$-boson
term [see Eq.~(\ref{floweq_Sg_sg})], that is,
\begin{equation}
 \frac{d}{d\Lam} Z_{\sg} \sim 
 - \frac{1}{4} \partial_{p_x}^2 \, [U(p)]^2 \alf^2
 \int_q D_{\Lam} \left. 
 \left[ G_{\pi\pi}(q) G_{\pi\pi}(p+q) \right] \, \right|_{p=0}
\end{equation}
for small $\Lam$.
$U(p)$ scales as $\Lam^{3-d}$, hence
$\frac{d}{d\Lam} Z_{\sg}$ scales as 
$\Lam^{-2} \Lam^{2(3-d)} \Lam^{d-4} = \Lam^{-d}$ for $1 < d < 3$,
so that $Z_{\sg}$ diverges as $\Lam^{1-d}$.
In three dimensions $U(p)$ vanishes only logarithmically for 
$\Lam \to 0$, so that $\frac{d}{d\Lam} Z_{\sg}$ scales as 
$|\log\Lam|^{-2} \Lam^{-3}$ and $Z_{\sg}$ diverges as 
$|\Lam \log\Lam|^{-2}$.
The scaling behavior of $m_{\sg}^2$ and $Z_{\sg}$ is not altered 
by the $\sg$-$\pi$ mixing term and the momentum dependence of 
the bosonic potential and therefore agrees with previous results
obtained for a simpler trunctation where $W=Y=0$.\cite{strack08}
The relation (\ref{rel_ZY}) implies that $Y$ is proportional
to $Z_{\sg}$ for $\Lam \to 0$, provided that $Z_{\pi}$ remains
finite as expected (and shown below). Hence, $Y$ also diverges
as $\Lam^{1-d}$ for $1 < d < 3$, and as $|\Lam \log\Lam|^{-2}$
in three dimensions.

The vanishing $\sg$-mass leads to divergent prefactors in the
flow equation (\ref{floweq_alf}) for the order parameter 
$\alf$. However, in the bosonic term this is compensated by the
bosonic potential, which scales exactly as $m_{\sg}^2 \,$, and the
fermionic integral vanishes linearly in $\Lam$, so that the flow
of $\alf$ always saturates at a finite value for $\Lam \to 0$.
One would run into artificially diverging flows for $\alf$ if
the fermionic cutoff were lowered too slowly compared to the 
bosonic one.\cite{strack08}
It is easy to see that the flows for $\Delta$ and $g_{\pi}$ 
also saturate at finite values for $\Lam \to 0$. 
The singularities of the bosonic contributions on the right hand
side of the corresponding flow equations are integrable.

We now show that the flow of $Z_{\pi}$ saturates at a finite
value for $\Lam \to 0$, so that the Goldstone mode maintains
a finite spectral weight, as expected.
This is not obvious, since the right hand side of the flow 
equation for $Z_{\pi}$ contains divergent terms.
From the flow equation (\ref{floweq_Sg_pi}) one obtains
\begin{eqnarray} \label{floweq_Z_pi}
 \frac{d}{d\Lam} Z_{\pi} &=&
 \frac{1}{2} \, \partial_{p_x}^2 \int_q \left.
 U(p+q) \, G'_{\pi\pi}(q) \, \right|_{p=0}
 \nonumber \\
 &-& \frac{1}{2} \, \partial_{p_x}^2 \int_q \left.
 \left[ U(q) \right]^2 \alf^2 D_{\Lam}
 \left[ G_{\sg\sg}(q) G_{\pi\pi}(p+q) \right] \, 
 \right|_{p=0} 
 \nonumber \\[2mm]
 &+& \mbox{finite terms} \; .
\end{eqnarray}
The first and the second term both diverge as $\Lam^{-1}$ for 
$\Lam \to 0$ for $1 < d < 3$, which suggests a logarithmic 
divergence of $Z_{\pi}$. However, these divergences cancel
each other.
To see this, we use the relation (\ref{rel_gamU}) to replace
the factor $\alf^2 U(q)$ in the second term by 
$\gam_{\sg\sg}(q) - \gam_{\pi\pi}(q)$.
For $\Lam \to 0$ one can neglect the $W$-term in the denominator 
of $G_{\sg\sg}$, such that 
$\gam_{\sg\sg}(q) \, G_{\sg\sg}(q) \to 1$, yielding
\begin{eqnarray}
 \frac{d}{d\Lam} Z_{\pi} &\sim&
 \frac{1}{2} \, \partial_{p_x}^2 \int_q \left.
 U(p+q) \, G'_{\pi\pi}(q) \, \right|_{p=0}
 \nonumber \\
 &-& \frac{1}{2} \, \partial_{p_x}^2 \int_q
 U(q) D_{\Lam} \big[ G_{\pi\pi}(p+q) 
 \nonumber \\
 && - \left. G_{\sg\sg}(q) \gam_{\pi\pi}(q) G_{\pi\pi}(p+q) \big] \, 
 \right|_{p=0} 
 \nonumber \\[2mm]
 &+& \mbox{finite terms} \; .
\end{eqnarray}
The divergent terms involving $G'_{\pi\pi}$ (and no other
propagator) cancel, and the remaining terms are finite for
$\Lam \to 0$, so that $Z_{\pi}$ indeed saturates.
Note that the momentum dependence of $U(q)$, parametrized by 
the naively irrelevant coupling $Y$, is crucial here.
Otherwise the first term in Eq.~(\ref{floweq_Z_pi}) would
vanish, and the divergence of the second term would remain
uncancelled. 

We finally derive the asymptotic behavior of the $\sg$-$\pi$
mixing parameter $W$. 
From the flow equation (\ref{floweq_Sg_sgpi}) for 
$\Sg_{\sg\pi}$, the last term yields the most singular
contribution to the flow of $W$,
\begin{equation}
 \frac{d}{d\Lam} W \sim 
 \partial_{p_0} \int_q
 \left. U(p) U(p-q) \alf^2 D_{\Lam} \left[
 G_{\sg\pi}(p-q) G_{\pi\pi}(q) \right] \, \right|_{p=0} \, ,
\end{equation}
where we have shifted the integration variable so that
the $p$-dependence appears in the argument of $G_{\sg\pi}$
instead of $G_{\pi\pi}$.
The factor $U(p)$ can be replaced by $U(0) = u$, since
$\left. \partial_{p_0} U(p) \right|_{p=0} = 0$.
Anticipating that $W$ vanishes sufficiently rapidly for
$\Lam \to 0$, we neglect the term of order $W^2$ in the
denominator of $G_{\sg\pi}(p-q)$.
Inserting the relation 
$\alf^2 U(p-q) = \gam_{\sg\sg}(p-q) - \gam_{\pi\pi}(p-q)$
then yields
\begin{eqnarray}
 && \frac{d}{d\Lam} W \sim 
 \nonumber \\
 && - u \, \partial_{p_0} \int_q
 \left. D_{\Lam} \left\{ \left[ 
 \frac{W(p_0 - q_0)}{\gam_{\pi\pi}(p-q)} - 
 \frac{W(p_0 - q_0)}{\gam_{\sg\sg}(p-q)} \right]
  G_{\pi\pi}(q) \right\} \, \right|_{p=0} ,
 \nonumber \\
\end{eqnarray}
where $G_{\pi\pi}(q) \sim \gam_{\pi\pi}^{-1}(q)$.
The second term in the bracket is subleading and can be
neglected.
Carrying out the scale derivative $D_{\Lam}$, and 
substituting $q \mapsto p - q$ in one of the resulting 
terms, the contributions can be combined to
\begin{equation}
 \frac{d}{d\Lam} W \sim 
 u \, \partial_{p_0} \int_q \left.
 \frac{dR_b(q)}{d\Lam} \,
 \frac{W p_0}{\gam_{\pi\pi}^2(q) \gam_{\pi\pi}(p-q)} \,
 \right|_{p=0} \; .
\end{equation}
Only the $p_0$ dependence in the numerator yields a
contribution to the frequency derivative at $p_0 = 0$. 
Using again $\gam_{\pi\pi}^{-1}(q) \sim G_{\pi\pi}(q)$,
one thus obtains the simple asymptotic flow equation
\begin{equation}
 \frac{d}{d\Lam} W \sim 
 - u W \int_q G_{\pi\pi}(q) \, G'_{\pi\pi}(q) \; .
\end{equation}
Comparing this with the asymptotic flow equation 
(\ref{asymp_m_sg}) for $m_{\sg}^2 \,$, we see that
$W^{-1} \frac{d}{d\Lam} W \sim
 (m_{\sg}^2)^{-1} \frac{d}{d\Lam} m_{\sg}^2 \,$, so that
\begin{equation}
 \frac{W}{m_{\sg}^2} \to C = \mbox{const}
\end{equation}
for $\Lam \to 0$. In other words, $W$ vanishes with the same
power of $\Lam$ as $m_{\sg}^2 \,$. This confirms that the term 
$W^2q_0^2$ is indeed subleading in the denominator of the 
bosonic propagators.
In an interacting Bose gas, the asymptotic ratio $C$ is 
proportional to the ''condensate compressibility'' 
$d\alf^2/d\mu$, where $\mu$ is the chemical potential for
the bosons.\cite{pistolesi04}

Since $Z_{\pi}$ and $W/m_{\sg}^2$ tend to finite constants
for $\Lam \to 0$, the progagators containing a $\pi$-field
assume the simple asymptotic form
\begin{eqnarray}
 G_{\pi\pi}(q) &\sim& 
 \frac{1}{Z_{\pi} (q_0^2 + \om_{\bq}^2)}
 \; , \\
 G_{\pi\sg}(q) &\sim& 
 \frac{Cq_0}{Z_{\pi} (q_0^2 + \om_{\bq}^2)} \sim 
 - G_{\sg\pi}(q)
\end{eqnarray}
for small $q$, in agreement with the behavior known for the 
interacting Bose gas.\cite{pistolesi04}
Only $G_{\sg\sg}(q)$ exhibits anomalous scaling, with a 
power-law depending on the dimensionality for $d < 3$, 
and logarithmic corrections in three dimensions.

%%%%%%%%%%%%%%%%%%%%%%%%%%%%%%%%%%%%%%%%%%%%%%%%%%%%%%%%%%%%%%%%%%%%%%%

\subsection{Flows in two dimensions}

We now present numerical results for the renormalization group 
flow of the variables parametrizing the effective action, using
the attractive two-dimensional Hubbard model as a prototypical
example.\cite{friederich10}
Thereby, the asymptotic behavior derived analytically 
in the preceding section is confirmed, and further interesting 
features at low and intermediate scales are obtained.
We choose a moderate attraction $U = -4t$, and fix the fermion
density at quarter-filling, where the Fermi surface is nearly 
circular and far from the Van Hove points in the Brillouin zone. 
\cite{fn3} All results are presented in units of $t$.

\begin{figure}[tb]
\centerline{\includegraphics[width=8cm]{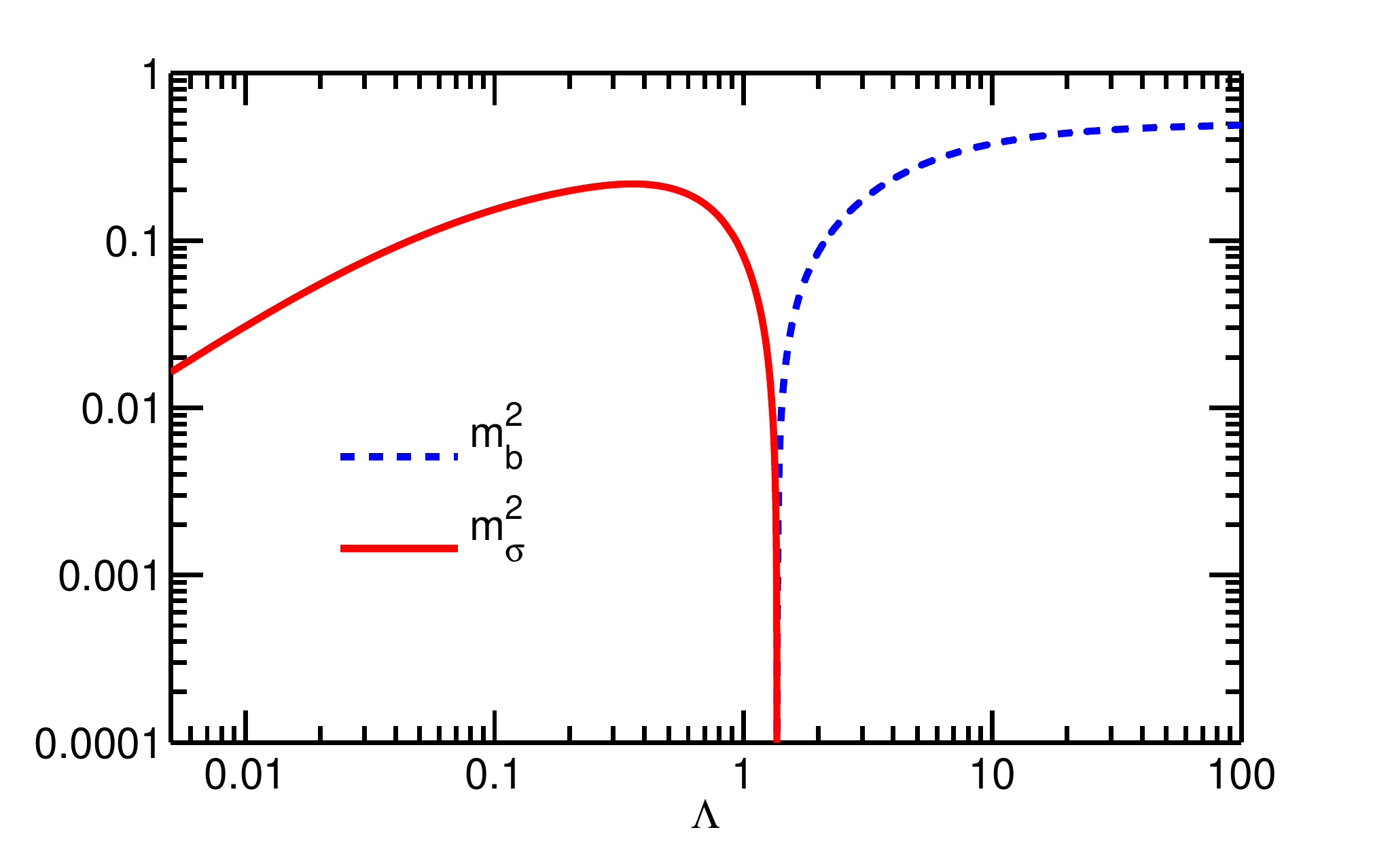}}
\caption{(Color online) Flow of the bosonic mass $m_b$ above 
 the critical scale $\Lam_c$, and the $\sg$-field mass 
 $m_{\sg}$ below $\Lam_c$.}
\end{figure}
The flow of the bosonic masses, $m_b$ for $\Lam$ above $\Lam_c$ 
and $m_{\sg}$ below, is shown in Fig.~7. 
The masses vanish at the critical scale $\Lam_c = 1.36$, and
$m_{\sg}^2$ decreases linearly in $\Lam$ for $\Lam \to 0$, in 
agreement with the analytic asymptotic results.

The flows of the gap $\Delta$ and the bosonic order parameter
$\alf$ are shown in Fig.~8. Both quantities saturate at finite
values for $\Lam \to 0$, where $\alf$ is slightly smaller than
$\Delta$. While $\Delta$ increases monotonically in the course
of the flow, $\alf$ is slightly decreased in the final stage 
by bosonic fluctuation contributions.
\begin{figure}[tb]
\centerline{\includegraphics[width=8cm]{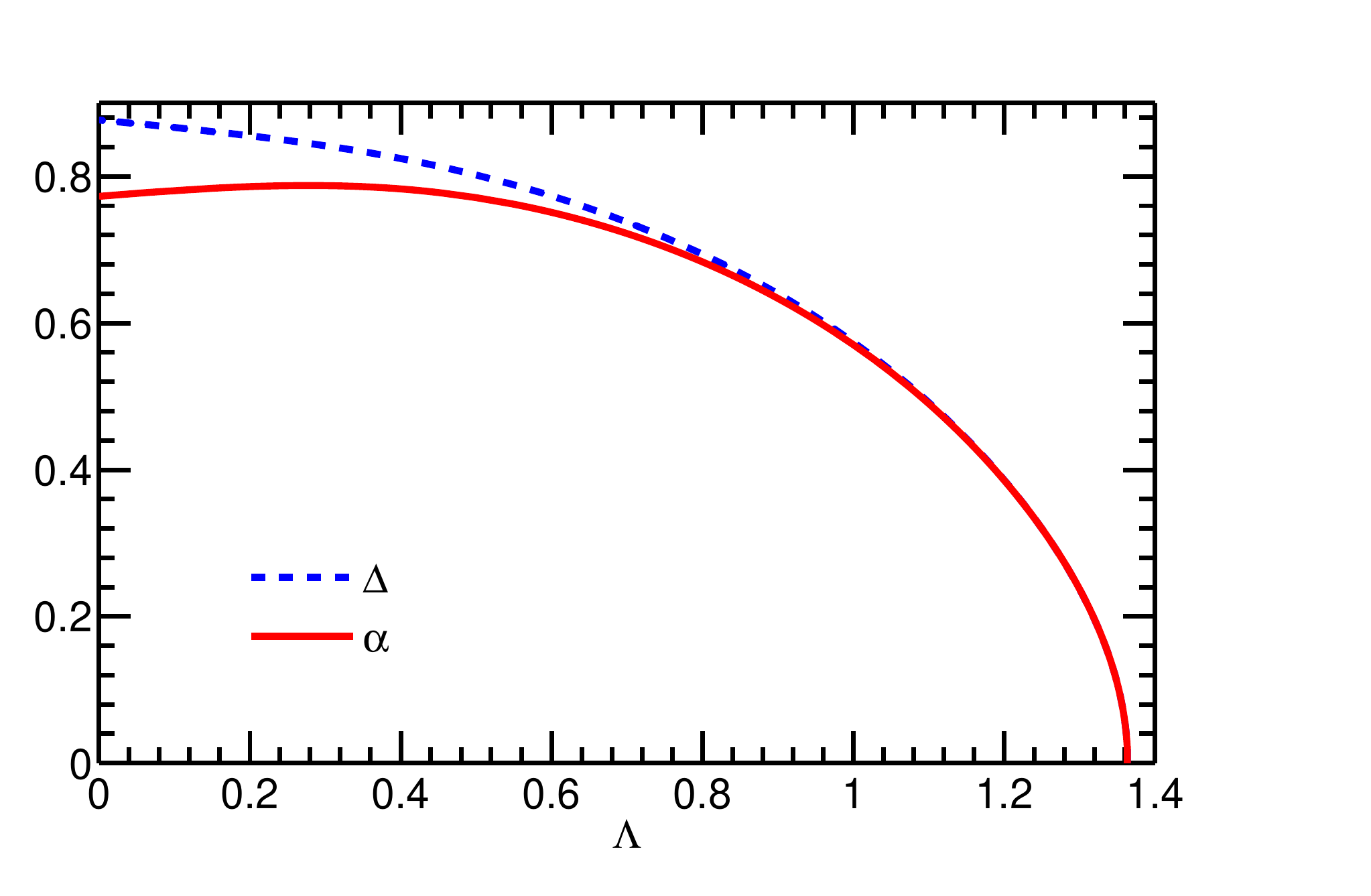}}
\caption{(Color online) Flows of the gap $\Delta$ and the 
 bosonic order parameter $\alf$ below the critical scale.}
\end{figure}
The final gap $\Delta = 0.88$ is reduced compared to the BCS 
mean-field gap $\Delta^{\rm MF} = 1.16$. The reduction is 
partially due to the truncation of the bosonic potential 
at quartic order,\cite{strack08} and, of course,
to fluctuations.
However, the reduction is weaker than expected, which can
be attributed to lacking fluctuation contributions from the 
fermionic particle-hole channel in our truncation.
Perturbation theory \cite{martin-rodero92} and fermionic 
RG calculations \cite{gersch08,eberlein13} yield a gap 
reduction to roughly one half.
The fermion-boson flow computed by Strack et al.\cite{strack08} 
yielded an even stronger gap reduction, caused mostly by a 
strong suppression of the critical scale by fluctuations 
in the symmetric regime.
A more accurate calculation of the gap within a bosonized fRG 
requires rebosonization of effective two-fermion interactions 
which are generated by the flow.\cite{floerchinger08}
We do not pay attention to these contributions, since our aim 
here is not an accurate estimate of the gap, but rather a 
comprehensive analysis of the low energy singularities.
\begin{figure}[tb]
\centerline{\includegraphics[width=8cm]{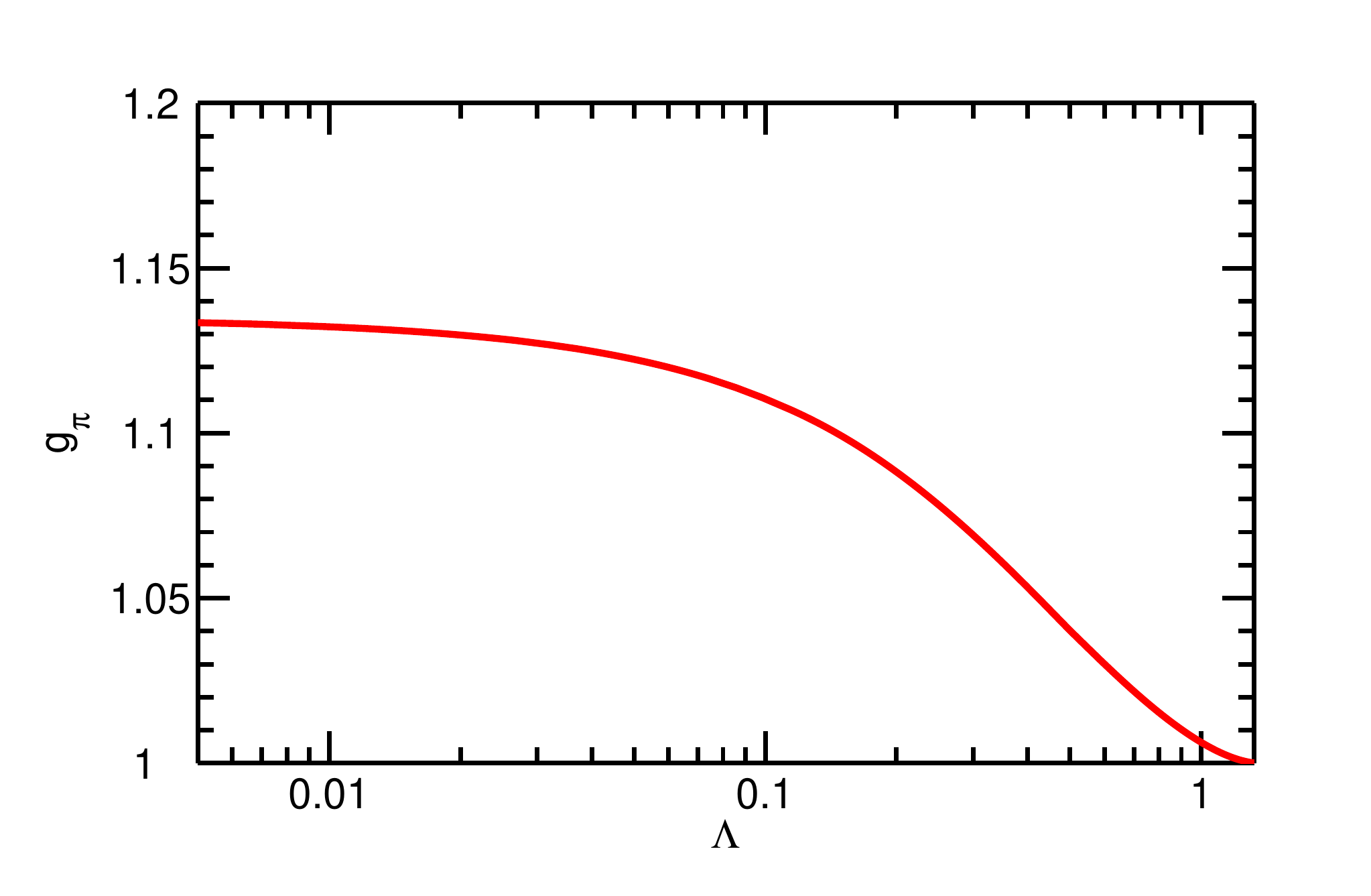}}
\caption{(Color online) Flow of the fermion-boson coupling 
 $g_{\pi}$ below the critical scale $\Lam_c$.}
\end{figure}
The flow of the fermion-boson coupling $g_{\pi}$ shown in
Fig.~9 is determined by the flow of $\Delta$ and $\alf$ via 
the Ward identity $\Delta = g_{\pi} \alf$. 
The increase of $g_{\pi}$ with decreasing $\Lam$ therefore 
reflects the increasing ratio $\Delta/\alf$ observed already 
in Fig.~8.

In Fig.~10 we show the flow of the prefactors of the quadratic
momentum and frequency dependence of the bosonic self-energies,
$Z_b$ above the critical scale, and $Z_{\sg}$, $Z_{\pi}$ below. 
\begin{figure}[tb]
\centerline{\includegraphics[width=8cm]{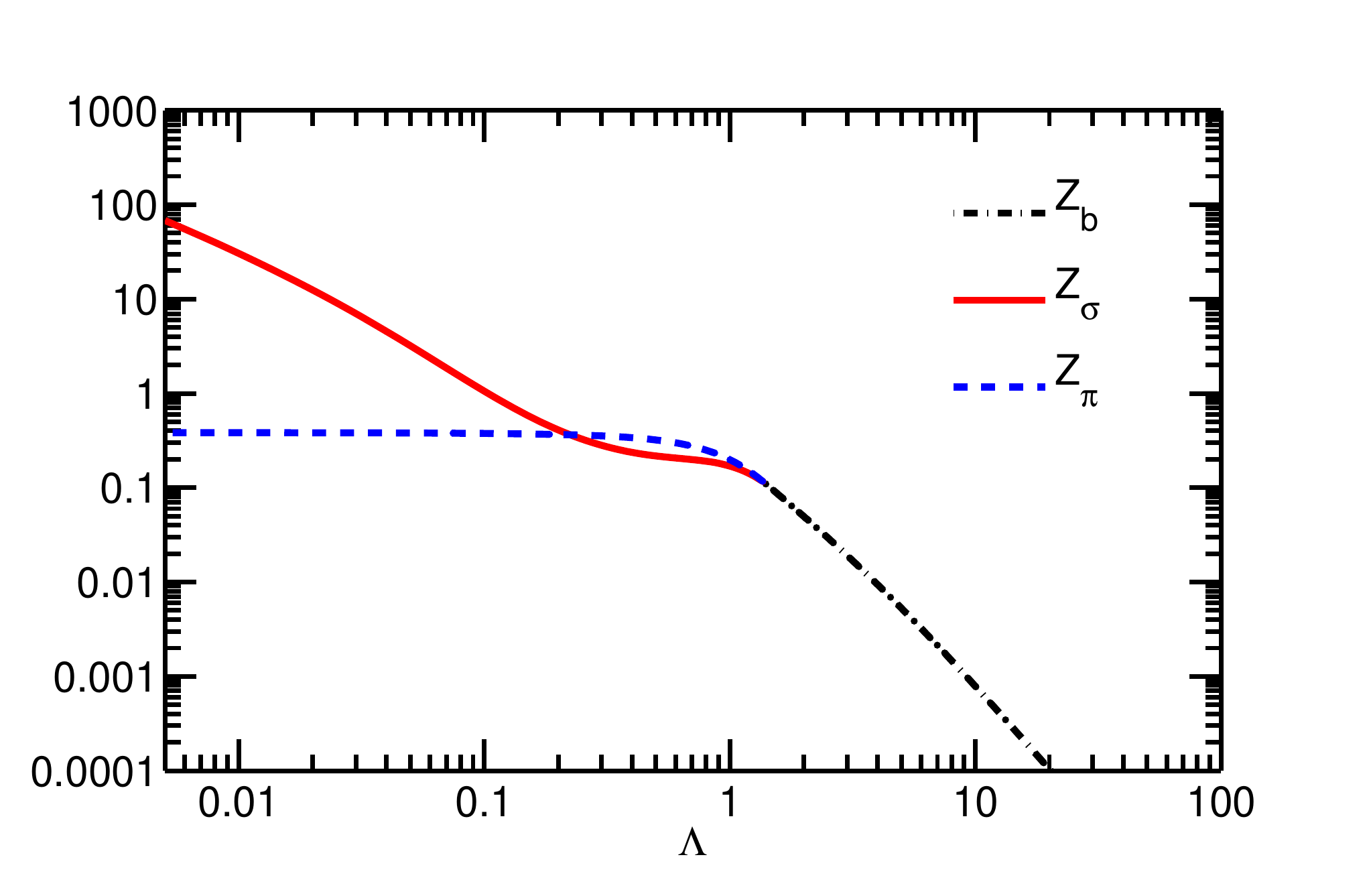}}
\caption{(Color online) Flow of $Z_b$ above the critical 
 scale $\Lam_c$, and of $Z_{\sg}$ and $Z_{\pi}$ below.}
\end{figure}
While $Z_{\pi}$ converges rapidly to a finite value, $Z_{\sg}$ 
first seems to saturate and falls even below $Z_{\pi}$, but 
then takes off to diverge as $\Lam^{-1}$ for small $\Lam$, in 
agreement with the analytic asymptotic results.

The flow of the bosonic interaction parameters $u$ and $Y$ is
displayed in Fig.~11. The local coupling $u$ first increases
upon lowering the scale, exhibits a weak kink at $\Lam_c$,
and finally decreases linearly in $\Lam$ for $\Lam \to 0$.
The latter bevavior is dictated by $m_{\sg}^2 \propto \Lam$ 
via the relation $u = m_{\sg}^2/\alf^2$. 
\begin{figure}[tb]
\centerline{\includegraphics[width=8cm]{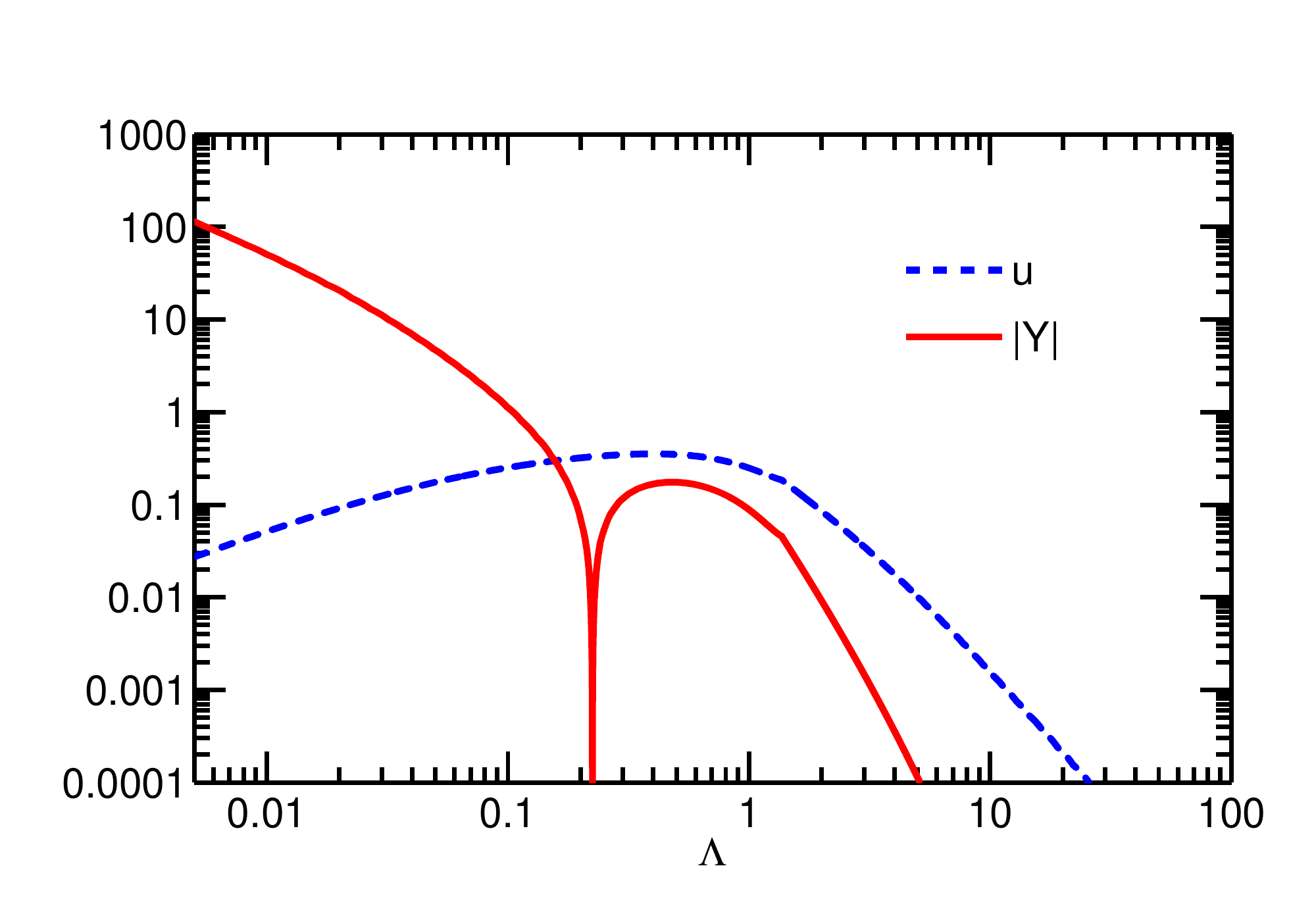}}
\caption{(Color online) Flow of the interaction parameters 
 $u$ and $Y$. The latter is initially negative and changes 
 sign after crossing zero at a scale situated an order of
 magnitude below $\Lam_c$.}
\end{figure}
The non-local coupling $Y$ is initially negative, and changes
sign at a scale $\Lam_*$ well below $\Lam_c$. 
The relation $\alf^2 Y = Z_{\sg} - Z_{\pi}$ thus implies that 
$Z_{\sg} < Z_{\pi}$ in the regime between $\Lam_*$ and 
$\Lam_c$, as is indeed observed in Fig.~10.
For small $\Lam$, the coupling $Y$ approaches $Z_{\sg}/\alf^2$
and therefore diverges as $\Lam^{-1}$.
%check Y in Figure, factor 4 too big !!!

Finally, in Fig.~12 we show the flow of the variable $W$
parametrizing the imaginary linear frequency contribution 
to the bosonic self-energy, which leads to a mixing of
$\sg$- and $\pi$-fields in the symmetry-broken regime.
\begin{figure}[tb]
\centerline{\includegraphics[width=8cm]{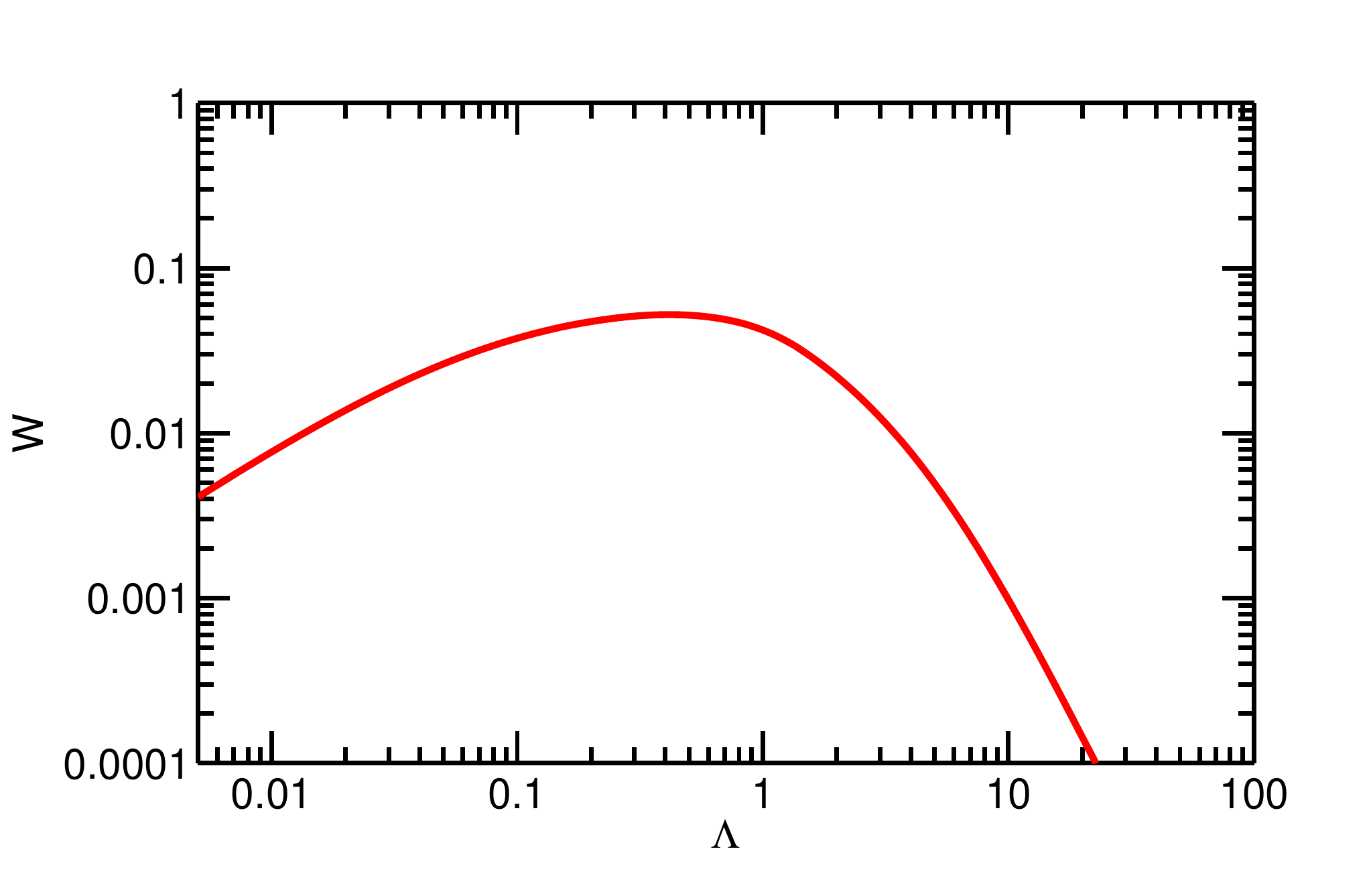}}
\caption{(Color online) Flow of the parameter $W$.}
\end{figure}
$W$ is always positive, increases until $\Lam$ has reached 
the critical scale and a bit beyond, and then decreases 
again with an asymptotic behavior proportional to $\Lam$ 
for $\Lam \to 0$, in agreement with the analytic analysis 
in the preceding section.
The ratio $W/m_{\sg}^2$ converges to $C = 0.25$ for 
$\Lam \to 0$.

%%%%%%%%%%%%%%%%%%%%%%%%%%%%%%%%%%%%%%%%%%%%%%%%%%%%%%%%%%%%%%%%%%%%

\section{Conclusion}

Charge-neutral fermionic superfluids exhibit rather complex
low-energy behavior due to singular renormalizations generated
by the Goldstone mode.
In particular, the mass for longitudinal order parameter
fluctuations scales to zero in dimensions $d \leq 3$.
We have found a relatively simple set of one-loop renormalization
group equations which fully captures these singularities, and 
conserves cancellations imposed by the $U(1)$ charge symmetry.
The flow equations are based on a symmetry-conserving truncation
of the fRG effective action.
Systematic cancellations guarantee that the flow conserves the 
massless Goldstone boson with a finite $Z$-factor, and that it 
respects an important Ward identity relating the pairing gap 
$\Delta$ to the bosonic order parameter and the fermion-boson 
vertex.
The truncation includes a self-energy term which mixes longitudinal 
and transverse order parameter fluctuations. 
We have shown that its renormalization is linked to the scaling 
of the longitudinal mass, in agreement with the behavior of an 
interacting Bose gas, where the mixing term is related to the 
condensate compressibility.\cite{pistolesi04}

Our truncation conserves the Goldstone theorem and the lowest
order Ward identity relating fermions and bosons.
The $U(1)$ symmetry actually entails an infinite hierarchy of 
Ward identities relating vertex functions of arbitrary order.
Any truncation of the effective action at finite order in the
fields leads to a deviation from the exact flow and a violation
of Ward identities involving vertex functions which are 
discarded in the truncation.
For example, there is a Ward identity relating the difference
between the longitudinal and transverse fermion-boson coupling,
$g_{\sg} - g_{\pi}$, to a two-fermion-two-boson vertex.
\cite{obert13}
If the latter is discarded, as in our truncation, one has to
set $g_{\sg} = g_{\pi}$ for the sake of consistency.

The main goal of this work was to deal with the singularities
associated with the Goldstone boson. 
We did not pay attention to finite renormalizations such as
contributions from the fermionic particle-hole channel or 
fermionic $Z$-factors.
However, building on the insights gained in the present work, 
one can construct improved truncations which yield more accurate 
results for the gap and other non-universal quantities, and 
still capture the singular low-energy behavior.

%%%%%%%%%%%%%%%%%%%%%%%%%%%%%%%%%%%%%%%%%%%%%%%%%%%%%%%%%%%%%%%%%

\begin{acknowledgments}
We thank A.~Eberlein, H.~Gies, N.~Hasselmann, A.~Katanin, 
J.~Pawlowski, M.~Salmhofer, P.~Strack and C.~Wetterich 
for useful discussions. 
\end{acknowledgments}

%%%%%%%%%%%%%%%%%%%%%%%%%%%%%%%%%%%%%%%%%%%%%%%%%%%%%%%%%%%%%%%%%%%%%

\begin{appendix}

\section{Ward identity}

Here we describe the derivation of the Ward identity,
Eq.~(\ref{WI}), relating the fermionic gap to the bosonic
order parameter and the fermion-boson vertex.
From the global $U(1)$ symmetry corresponding to charge 
conservation one can derive the following identity for the
effective action,\cite{fn4}
\begin{equation} \label{WI1}
 \int_{p\sg} \left( 
 \psi_{p\sg} \frac{\delta\Gam}{\delta\psi_{p\sg}} -
 \psib_{p\sg} \frac{\delta\Gam}{\delta\psib_{p\sg}} 
 \right) =
 2 \int_q \left(
 \phi_q^* \frac{\delta\Gam}{\delta\phi_q^*} -
 \phi_q \frac{\delta\Gam}{\delta\phi_q} \right) \; .
\end{equation}
This relation holds at any scale $\Lam$, since the 
regulator does not spoil the symmetry.
From this functional identity one can derive Ward 
identities for vertex functions by taking functional
derivatives with respect to the fields at $\psi_{p\sg} 
= \psib_{p\sg} = 0$ and $\phi_q = \alf \delta_{q0}$.
In particular, applying 
$\frac{\delta^2}{\delta\psib_{k\up}\delta\psib_{-k\down}}$
yields 
\begin{equation} \label{WI2}
 \frac{\delta^2 \Gam}
 {\delta\psib_{k\up}\delta\psib_{-k\down}} =
 \alf \frac{\delta^3 \Gam}
 {\delta\psib_{k\up}\delta\psib_{-k\down}\delta\phi_0} -
 \alf^* \frac{\delta^3 \Gam}
 {\delta\psib_{k\up}\delta\psib_{-k\down}\delta\phi_0^*}
 \; ,
\end{equation}
that is, a relation between the gap and the fermion-boson
vertices. Applying 
$\frac{\delta^2}{\delta\psi_{k\up}\delta\psi_{-k\down}}$
yields the complex conjugate relation.
For our ansatz, with a $k$-independent gap and a 
$k$-independent fermion-boson coupling, Eq.~(\ref{WI2}) 
yields
\begin{equation}
 \Delta = g \alf - \tilde g \alf^* \; .
\end{equation}
Choosing $\alf$ and $\Delta$ real, one obtains the Ward 
in the form of Eq.~(\ref{WI}).

\end{appendix}

%%%%%%%%%%%%%%%%%%%%%%%%%%%%%%%%%%%%%%%%%%%%%%%%%%%%%%%%%%%%%%%%%%%%


\begin{thebibliography}{99}
\bibitem{fn1} Note that in Ref.~\onlinecite{strack08} the bosonic 
 propagator was defined with a minus sign, while the prefactor 
 $\frac{1}{2}$ in front of the expectation value was absent.
\bibitem{fn2} Here and in the following flow equations involving
 momentum derivatives, we assume 
 $\lim_{|\bq| \to 0} \om_{\bq}^2/\bq^2 = 1$ to simplify the
 equations. For a general asymptotic ratio the right hand side
 of the flow equations have to be divided by
 $\lim_{|\bq| \to 0} \om_{\bq}^2/\bq^2$.
\bibitem{fn3} The corresponding chemical potential is $\mu = 
 -1.44t$. There is no interaction induced shift of the Fermi 
 surface since the normal fermion self-energy is not taken
 into account.
\end{thebibliography}
\end{document}